\documentclass[apj, twocolumn, twocolappendix, appendixfloats]{openjournal}
\usepackage{amsmath}
\usepackage{booktabs}
\usepackage{graphicx}
\usepackage[utf8]{inputenc}
\usepackage{hyperref}
\hypersetup{
    colorlinks=true,
    linkcolor=blue,
    filecolor=blue,
    urlcolor=blue,
    citecolor=blue,
}
\usepackage{orcidlink}
\usepackage{soul}
\usepackage{color, colortbl}
\usepackage[nolist]{acronym}
\usepackage{newtxtext, newtxmath, enumitem, multirow, ctable}
\usepackage[export]{adjustbox}

\usepackage{macros}

\graphicspath{{figure/}}

\begin{document}

\title{\vspace{-0.9 cm}
Radial redistribution of stellar orbits in FIRE simulations of Milky-Way-mass galaxies
\vspace{-1.6 cm}}

\shortauthors{Bellardini et al.}

\author{Matthew A. Bellardini\,\orcidlink{0000-0002-5663-207X}$^1$}
\author{Fiona McCluskey$^1$}
\author{Andrew Wetzel\,\orcidlink{0000-0003-0603-8942}$^1$}
\author{Cecilia Steel$^1$}
\author{Sarah R. Loebman\,\orcidlink{0000-0003-3217-5967}$^2$}
\author{Jorge Moreno\,\orcidlink{0000-0002-3430-3232}$^3$}

\thanks{Corresponding author: Andrew Wetzel}
\email{awetzel@ucdavis.edu}

\affiliation{$^1$ Department of Physics \& Astronomy, University of California, Davis, CA 95616, USA}
\affiliation{$^2$ Department of Physics, University of California, Merced, CA 95343, USA}
\affiliation{$^3$ Department of Physics \& Astronomy, Pomona College, Claremont, CA 91711, USA}

\begin{abstract}
A central question in galactic dynamics and galactic archeology is: how much do the orbits of stars redistribute (migrate) after birth?
We use the FIRE-2 cosmological zoom-in simulations of 11 Milky Way-mass galaxies to quantify the change in the orbital specific angular momentum, $\jphi$, orbital radius, $\Rorbit$, and azimuthal velocity, $\vphi$, of stars from birth to today.
We examine the dependences on stellar age, present-day $\Rorbit$, and birth $\Rorbit$, characterizing both the median (net) change, $\dRorbit$, and its scatter, $\sigma(\dRorbit)$.
We comprehensively compare five ways of measuring orbital radius; we find generally consistent trends, but only when measuring radius today and radial redistribution self-consistently.
Stars selected by their \textit{birth} $\Rorbit$ typically decreased in $\Rorbit$, $\jphi$, and $\vphi$ since birth.
The trend for stars at a given $\Rorbit$ \textit{today} depends on age: those younger than $\approx5\Gyr$ generally decreased in $\Rorbit$, $\jphi$, and $\vphi$ since birth, while those older generally increased in $\Rorbit$, $\jphi$, and $\vphi$ since birth.
$\sigma(\dRorbit)$, a standard metric of radial redistribution, increases with stellar age only up to $\approx3\Gyr$; it saturates at $\sigma(\dRorbit)\approx2\kpc$ for older stars.
This saturation contradicts a common expectation of a monotonic increase with age.
Our results broadly agree with recent observational inferences of $\dRorbit$ and $\sigma(\dRorbit)$ in the Milky Way.
Across our FIRE-2 sample, the timing of disk formation does not correlate with $\sigma(\dRorbit)$, but it correlates with (net) $\dRorbit$.
\end{abstract}

\section{Introduction}

The stars within a galaxy provide a record of its formation over cosmic time, including its cosmological assembly and its dynamical evolution.
A key goal of galactic archeology of the Milky Way (MW) is to use the observed positions, velocities, elemental abundances, and ages of its stars today to reveal its history and understand the physical processes underlying galaxy formation \citep[for example][]{FreemanBlandHawthorn2002, RixBovy2013, Deason2024}.
The advent of \textit{Gaia} \citep{Gaia2016}, along with spectroscopic surveys such as APOGEE \citep{Majewski2017}, LAMOST \citep{Cui2012}, GALAH \citep{DeSilva2015, Buder2018}, and now SDSS-V Milky Way Mapper \citep{SDSS2025}, have ushered in a golden age of MW studies, by measuring millions-to-billions of stars across the MW.
However, connecting the MW's present state to its past remains challenging, in large part because of `radial redistribution': stars currently may reside at galactocentric radii, $R$, far from where they formed.
Such changes to the orbital $R$ of a population of stars scrambles our record of a galaxy's history, in particular, of its star-forming interstellar medium (ISM), which sets the conditions of stars at birth.
As we discuss below, many studies have sought to understand the radial redistribution of stars, both as a dynamical process in its own right, and to reverse its effects on the spatial and velocity distributions of stars today to uncover stars' birth conditions and infer the formation history of the MW.

We commonly describe the orbits of stars in terms of (1) specific angular momentum (or azimuthal action), $j_{\phi}$, which fixes the `guiding center radius', $R_{j}$, which is the radius of a circular orbit with the same $j_{\phi}$, and (2) radial and vertical actions, $j_{R}$ and $j_{Z}$.
The latter two set the amplitude of epicyclic excursions around $R_{j}$ and perpendicular to the disk plane, respectively.
Radial excursions (from eccentric orbits) alone will cause a star’s \textit{instantaneous} radius, $R_{\rm phys}$, to vary throughout its orbit, even absent dynamical evolution of its orbit.
However, more important are changes to the actual orbits of stars.

One can divide radial redistribution into changes in random energy (or eccentricity) and changes in $j_{\phi}$ (or $R_{j}$).
During their lifetimes, stars undergo repeated scattering interactions with perturbations within the disk, such as giant molecular clouds \citep[GMCs;][]{SpitzerSchwarzschild1951, Wielen1977, Lacey1984}, spiral arms \citep{BarbanisWoltjer1967, CarlbergSellwood1985, MinchevQuillen2006}, and bars \citep{Saha2010, Grand2016a}, and outside the disk, such as satellite or merging galaxies \citep{Quillen2009, Bird2012, Carr2022}. 
These interactions can increase the random energy of a star's orbit and/or alter $j_{\phi}$.
\citet{SellwoodBinney2002} showed that scattering at the corotation resonance of transient spiral arms could drive significant changes in $j_{\phi}$ \textit{without} increasing random energy.
This mechanism implies that radial redistribution might not leave a dynamical signature in terms of an increase in random energy (eccentricity).
Later works found that resonant overlap, be it spiral-spiral or bar-spiral \citep{MinchevFamaey2010, Minchev2012b}, and the formation and subsequent slowing of the bar \citep{Chiba2021, Haywood2024, Khoperskov2024} can drive even more substantial changes in $j_{\phi}$, and in certain cases, maintain or even increase orbital circularity.

To clarify our terminology, throughout we refer to \textit{any} changes to galactocentric radius, $R$, as `radial redistribution'.
In the literature, many works refer to `radial migration', as \citet{SellwoodBinney2002} introduced, but some works use it as a general term for any radial displacement, others use it as a shorthand for shifts in $R_{j}$ that broadly conserve circularity, and others use it to refer to the specific case of corotation scattering off of spiral-arms.
We use `cold torquing' \citep{Daniel2019} to refer to changes in $R_{j}$ that broadly conserve circularity (do not significantly change $j_{R}$ or $j_{Z}$), a scenario also called `churning' \citep{SchoenrichBinney2009a}.
We refer to `heating' as changes in radial and/or vertical action, $j_{R}$ and/or $j_{Z}$, a scenario also called `blurring'.
(Though, a change to radial action also could `cool' an originally non-circular orbit.)
Of course, many dynamical perturbations can produce both torquing and heating at the same time.

In an ideal scenario, one can approximate cold torquing and heating as diffusion processes.
Here, a population of stars born at the same time and $R$ can experience perturbations that drive both inward and outward radial redistribution, broadening the resultant distribution about its birth $R$ \citep{Wielen1977, Brunetti2011, Sellwood2014}.
However, simulations and analytic studies show that stellar populations also can undergo coherent net shifts inward or outward \citep{Halle2015, ElBadry2016, Khoperskov2020}.
Importantly, in this work, we examine both the diffusive and coherent (net) aspects of radial redistribution.

Stellar abundances provide long-lived birth records: in general, the metallicity of the ISM, and of stars at birth, increases over time within a galaxy, and many disk galaxies have negative radial gradients in metallicity.
Therefore, in principle, within a disk galaxy without radial redistribution, one would expect fairly tight relations between metallicity, age, and $R$.
Instead, the solar neighborhood shows little correlation between metallicity and age \citep{Edvardsson1993}, and also contains relatively old stars with super-solar metallicities \citep{Grenon1989, Castro1997}.
The cold (near circular) orbits observed for most of these stars, including the old metal-rich populations, limit the extent of their radial excursions.
\citet{SellwoodBinney2002} showed that cold torquing could address this tension between galactic elemental evolution models and observations of the solar neighborhood, specifically, if these stars changed $R$ without substantially increasing their radial actions.
Subsequent work confirmed the presence of cold torquing in baryonic (including hydrodynamics) simulations \citep{Roskar2008b}, and its ability to account for the dispersion in the relation between stellar age and metallicity \citep[for example][]{Haywood2008, Casagrande2011, Carrillo2023}, supersolar metallicity stars in the solar neighborhood \citep[][]{Kordopatis2015, Anders2017, Feuillet2018}, and the skewness of the metallicity distribution as a function of $R$ \citep[for example][]{Hayden2015, Loebman2016, MartinezMedina2016}.

Nevertheless, the literature shows little consensus on the strength of radial redistribution, its dependence on age, and its impact on galactic properties today.
Some works claim that radial redistribution is a major driver of various aspects of the MW's structure \citep{SchoenrichBinney2009b, Hayden2015, Ratcliffe2023}, including the local prominence of the bimodality in $\alpha$ elements \citep{Sharma2021alpha}, while others find that it likely induces only slight changes around formation driven trends \citep{Spitoni2015, Haywood2019, Khoperskov2021, Yu2023, Graf2025a}.

This lack of consensus is in large part because almost all inferences of the strength of radial redistribution in the MW require significant assumptions about the properties of the MW in its past.
A galaxy's present-day structure -- in terms of the dynamics, ages, and metallicities of its stars -- is the convolution of two effects: the state of its star-forming ISM over time and the subsequent dynamical evolution of its stars.
Because we do not know either one independently, most works impose assumptions about the past ISM conditions, for example, that the metallicity always declined with $R$ and/or with minimal azimuthal variations, the past star formation history, and the time and spatial dependences of radial redistribution \citep{Frankel2018, Ratcliffe2023, Zhang2025bimodal}.
However, different assumptions can produce dramatically different conclusions. 
For example, \citet{SchoenrichBinney2009a} highlighted the degeneracy between the strength of the ISM metallicity radial gradient and the strength of cold torquing: the steeper the metallicity gradient in the ISM, the smaller the effect of cold torquing. 
Implicit in many models is that the MW's present state broadly represents its earlier state, for example, that stars formed in (near) circular orbits within a dynamically cold (thin) disk throughout much/all of the MW's history, and that scattering interactions were primarily driven by spiral arms, bars, and GMCs, with properties broadly similar to their current values \citep[for example][]{Minchev2018, Frankel2020, Sharma2021dynamics}.

In contrast, observations of high-redshift galaxies indicate that the progenitors of MW-mass galaxies were thicker, clumpier, and more turbulent than their descendants today \citep[for example][]{Kassin2012, Glazebrook2013, Conselice2014, Ubler2019, Wisnioski2019, Danhaive2025}, with higher gas fractions, higher (and burstier) star formation rates, more frequent mergers \citep[for example][]{Tacconi2013, Tacconi2020, MadauDickinson2014, FoersterSchreiberWuyts2020}, and nearly flat radial gradients in ISM metallicity \citep[for example][]{Wang2022, Ju2025, Sun2025}.
Thus, thin, well-ordered disks as observed in the local Universe do not necessarily represent the states of galaxies earlier in their history.
Furthermore, while many previous works considered dynamical scattering via internal structures within a disk, later works highlighted the role of external perturbers.
Simulations show that encounters with satellites and minor mergers warp and flare disks \citep[for example][]{SotilloRamos2023}, while also driving strong redistribution: the median amount of radial redistribution nearly doubles when disks are perturbed by orbiting satellites \citep{Bird2012}, while minor mergers can induce shifts of $3 - 5 \kpc$ \citep{VillalobosHelmi2008, Kazantzidis2008}.

Indeed, much theoretical work indicates that radial redistribution likely operated differently at earlier cosmic times.
For one, the efficiency of dynamical heating, and, to an even greater degree, cold torquing, depends on the dynamical state of a stellar population.
Dynamically cold populations couple strongly to nonaxisymmetric patterns and readily experience cold torquing, whereas thick, dynamically hot populations are much less responsive to such resonant perturbations \citep{VeraCiro2014, Daniel2015}.
In turn, the hotter dynamics of gas, and thus newly formed stars, observed at higher redshifts may imply weaker radial redistribution.
Similarly, lower-mass/higher-redshift galaxies typically hosted weaker spiral arms and bars, if at all \citep[for example][]{Melvin2014}, meaning the main drivers of radial redistribution in present-day MW-mass galaxies may be less effective or even absent at earlier times. 
However, other processes that drive radial redistribution may be more prevalent at early times.
For example, in cosmological simulations of low-mass galaxies, bursty star formation and its associated feedback drive rapid fluctuations in the gravitational potential that displace stars by several kiloparsecs over repeated cycles and build diffuse, extended stellar distributions \citep{ElBadry2017, Wheeler2019}. 

Therefore, interpreting the present-day record of the MW and MW-like galaxies requires models that incorporate their cosmological evolution, including the formation and settling of a disk, along with the associated changes in gas accretion, star formation, and mergers.
Cosmological zoom-in simulations are well-suited tools for understanding the strength of radial redistribution of stars, because they have sufficient resolution to self-consistently model the evolution of the clumpy, multi-phase ISM and its evolving dynamics over cosmic time, while also resolving individual star-forming regions and the sources of dynamical perturbations, both internal and external, that drive radial redistribution, all within a fully cosmological context \citep{Grand2017, Okalidis2022}.

We use FIRE-2 simulations of 11 MW-mass galaxies to examine the strength of stellar radial redistribution, from birth until today, in terms of changes to orbital radius, orbital velocity, and specific angular momentum.
In particular, we quantify how radial redistribution, both its net inward/outward motion, $\Delta R$, and its scatter, $\sigma(\Delta R)$, depend on stellar age and $R$, both at birth and today.
We examine how these results depend on the definition of orbital radius used.

\section{Methods}
\label{sec:methods}

\subsection{FIRE-2 simulations}

We use cosmological zoom-in simulations from the Feedback In Realistic Environments (FIRE) project \citep{Hopkins2018b}\footnote{
FIRE project web site: \href{http://fire.northwestern.edu}{http://fire.northwestern.edu}
}, all of which are publicly available \citep{Wetzel2023}.
We include $5$ cosmologically isolated MW/M31-mass galaxies from the \textit{Latte} suite \citep[introduced in][]{Wetzel2016} and $6$ Local Group-like MW$+$M31 pairs from the `ELVIS on FIRE' suite \citep{GarrisonKimmel2019a, GarrisonKimmel2019b}, which have halo masses $\Mthm = 1 - 2 \times 10^{12} \Msun$, where $M_{\rm 200m}$ is the total mass within the radius within which the mean density is $200$ times the mean matter density of the Universe.
Following \citet{Bellardini2021} and \citet{Bellardini2022}, we exclude 3 galaxies from the \textit{Latte} suite: m12r and m12z, because of their low stellar masses, and m12w, because of its unusually compact disk.
Table~\ref{table:galaxies} lists the galaxies that we analyze, along with key properties today.

The initial conditions of the simulations are embedded inside periodic cosmological boxes with side length $70.4 - 172 \Mpc$.
We generated them at $z \approx 99$ using \textsc{MUSIC} \citep{HahnAbel2011}.
Each simulation saved 600 snapshots to $z = 0$, with a typical time spacing of $\approx 25 \Myr$.
Each simulation uses a flat $\Lambda$CDM cosmology with parameters generally consistent with \citet{Planck2020b}: $h = 0.68 - 0.71$, $\Omega_{\Lambda} = 0.69 - 0.734$, $\Omega_{\rm m} = 0.266 - 0.31$, $\Omega_{\rm b} = 0.0455 - 0.048$, $\sigma_{\rm 8} = 0.801 - 0.82$ and $n_{\rm s} = 0.961 - 0.97$.

The initial baryon particle mass in the \textit{Latte} suite is $7070 \Msun$ ($\approx 5000 \Msun$ today because of stellar mass loss), and the dark-matter mass resolution is $3.5 \times 10^5 \Msun$.
Mass resolution is $\approx 2 \times$ higher in the ELVIS simulations: Thelma \& Louise has initial baryon particle masses of $4000 \Msun$ and Romeo \& Juliet and Romulus \& Remus have $3500 \Msun$.
The gravitational force softening of star and dark-matter particles is fixed, with a Plummer equivalent of $\epsilon_{\rm star} = 4 \pc$ and $\epsilon_{\rm dm} = 40 \pc$ (comoving at $z > 9$, physical thereafter).
Force softening for gas cells is adaptive, matching the hydrodynamic kernel smoothing, reaching a $1 \pc$ minimum; the value in the typical interstellar medium (ISM), at densities $\approx 1 \cci$, is $\approx 40 \pc$.

We ran all simulations with the Meshless Finite Mass hydrodynamics method \citep[][]{Hopkins2015} and with the FIRE-2 model for star formation and stellar feedback \citep{Hopkins2018b}.
Gas cells experience metallicity-dependent radiative heating and cooling processes across a temperature range of $10 - 10^{10}$K, including free-free, photoionization and recombination, Compton, photo-electric and dust collisional, cosmic ray, molecular, metal-line, and fine structure processes, accounting for $11$ elements (H, He, C, N, O, Ne, Mg, Si, S, Ca, Fe).
The simulations include a spatially-uniform time-dependent cosmic ultraviolet background from \citet{FaucherGiguere2009}.

Star formation occurs only in gas that is self-gravitating, Jeans-unstable, cold ($T < 10^4$K), and molecular \citep[following][]{KrumholzGnedin2011}.
Once a gas cell becomes eligible for star formation, it converts to a star particle on a local free-fall time.
A star particle inherits the mass of its progenitor gas cell and represents a single stellar population, assuming a \citet{Kroupa2001} initial mass function.
The FIRE-2 simulations model time-resolved stellar feedback processes such as continuous mass loss from stellar winds, core-collapse and white-dwarf (Type Ia) supernovae, radiation pressure, photoionization, and photoelectric heating.
They model stellar winds and their yields following a combination of models \citep{vandenHoek1997, Marigo2001, Izzard2004} synthesized in \citet{Wiersma2009}.
They use core-collapse supernova rates from \textsc{Starburst99} \citep{Leitherer1999} and white-dwarf supernova rates from \citet{Mannucci2006}.
They use \citet{Nomoto2006} and \citet{Iwamoto1999} for the nucleosynthetic yields of core-collapse and white-dwarf supernovae, respectively.
These simulations explicitly model unresolved turbulent diffusion of metals in gas \citep{Hopkins2017, Su2017, Escala2018}.

Crucial to this analysis, FIRE-2 simulations self-consistently model the multiphase ISM, including overdensities that can scatter the orbits of stars to drive radial redistribution.
For example, they resolve giant molecular clouds \citep{Benincasa2020, Guszejnov2020}, the formation of (massive) star clusters \citep{Ma2020a, Grudic2023, Sameie2023, Bhattarai2024}, spiral arms \citep{Orr2023, Quinn2025}, and bars \citep{Ansar2025}.
The emergence of these structures within the FIRE-2 galaxies, in addition to cosmological effects like gas accretion, galaxy mergers, and satellite interactions, together enables us to explore stellar radial redistribution in a self-consistent cosmological context without being beholden to any assumptions or analytic approximations regarding the evolution or impact of these structures.

\begin{table}
\centering
\caption{
properties today of the 11 FIRE-2 galaxies we analyze
}
\begin{tabular}{|c|ccc|}
\hline
galaxy & $M^\star_{\rm 90}$ & $R^\star_{\rm 90}$ & $t_{\rm lb} \left( v_{\phi} / \sigma_{v \rm,3D} > 1  \right)$ \\
name & [$10^{10} \Msun$] & [kpc] & [Gyr] \\
\hline
m12m$^{1}$ & 10.0 & 11.6 & 9.2 \\
Romulus$^{2}$ & 8.0 & 12.9 & 7.4 \\
m12b$^{3}$ & 7.3 & 9.0 & 7.4 \\
m12f$^{4}$ & 6.9 & 11.8 & 7.4 \\
Thelma$^{3}$ & 6.3 & 11.2 & 4.3 \\
Romeo$^{3}$ & 5.9 & 12.4 & 11.0 \\
m12i$^{5}$ & 5.3 & 9.8 & 6.7 \\
m12c$^{3}$ & 5.1 & 9.1 & 6.5 \\
Remus$^{2}$ & 4.0 & 11.0 & 7.9 \\
Juliet$^{3}$ & 3.3 & 8.1 & 4.4 \\
Louise$^{3}$ & 2.3 & 11.2 & 7.2 \\
\hline
\end{tabular}
\tablecomments{
Galaxies are in decreasing order of stellar mass today.
Columns show: galaxy name; $M^\star_{90}$ is the stellar mass within $R^\star_{90}$; $R^\star_{90}$ is the radius that encloses $90\%$ of the mass of all stars within $\pm 3 \kpc$ of the disk midplane; the lookback time to the onset of disk formation, defined in Table~1 of \citet{McCluskey2024} as when $v_{\phi} / \sigma_{v \rm,3D} > 1$ permanently for stars at birth.
The publication that introduced each simulation at this resolution is: \citet{Hopkins2018b}$^{1}$, \citet{GarrisonKimmel2019a}$^{2}$, \citet{GarrisonKimmel2019b}$^{3}$, \citet{GarrisonKimmel2017}$^{4}$, \citet{Wetzel2016}$^{5}$.
}
\label{table:galaxies}
\end{table}

\subsection{Eras of Disk Formation}
\label{sec:transition_epochs}

Various works quantified the formation histories of the MW-mass galaxies in these FIRE-2 simulations, including when a single primary progenitor galaxy emerged \citep{Santistevan2020, Horta2024}, disk formation and rotational motion \citep{Ma2017b, GarrisonKimmel2018, Yu2023}, the transition from bursty to smooth star formation \citep{Yu2021, Gurvich2023}, and their physical causes \citep{Hopkins2023b}.
Most relevant for this work, \citet{McCluskey2024} examined the orbital dynamics of stars across cosmic time and argued that these disks form and settle in three distinct eras: `pre-disk' (ages $\gtrsim 8 \Gyr$) when stars formed on largely dispersion-dominated orbits, `early-disk' (age $\approx 4 - 8 \Gyr$) when a (thick) disk emerged, dominated by net rotation but still dynamically hot ($v_{\phi} / \sigma_{v\rm,3D} > 1$), and `late-disk' (ages $\lesssim 4 \Gyr$) when stars formed on thin-disk, near-circular orbits (($v_{\phi} / \sigma_{v\rm,3D} \gtrsim 3$).
Throughout, we highlight trends for radial redistribution in these 3 eras, given that many trends show distinct behavior in each era.

\citet{McCluskey2024} also quantified the lookback time to the onset of disk formation for each galaxy, when $\vphi / \sigma_{v\rm,3D}$ of its stars at birth permanently rose above 1, which we list in Table~\ref{table:galaxies}.
This time represents when each galaxy `spun up' and transitioned from its pre-disk to early-disk disk era.

We are motivated in part by applications to the MW, and several observational analyses suggest that the MW's disk began to form unusually early, $\approx 12 \Gyr$ ago \citep[for example][]{BelokurovKravtsov2022, Conroy2022, Xiang2022}.
Therefore, in some figures we show results separately for Romeo, the galaxy in our sample whose disk began to form the earliest and likely closest in time to the MW.
As \citet{McCluskey2024} showed, Romeo began to spin up (in terms of $\vphi / \sigma_{v\rm,3D} \gtrsim 1$) $11 - 12 \Gyr$ ago, likely within $\approx 1 \Gyr$ of the MW.
This early formation is likely because Romeo formed in a Local Group-like environment \citep[see][]{GarrisonKimmel2018, Santistevan2020}.

\subsection{Measuring orbital radius}
\label{sec:radius_metric}

\begin{figure*}
\centering
\includegraphics[width = 0.99 \linewidth]{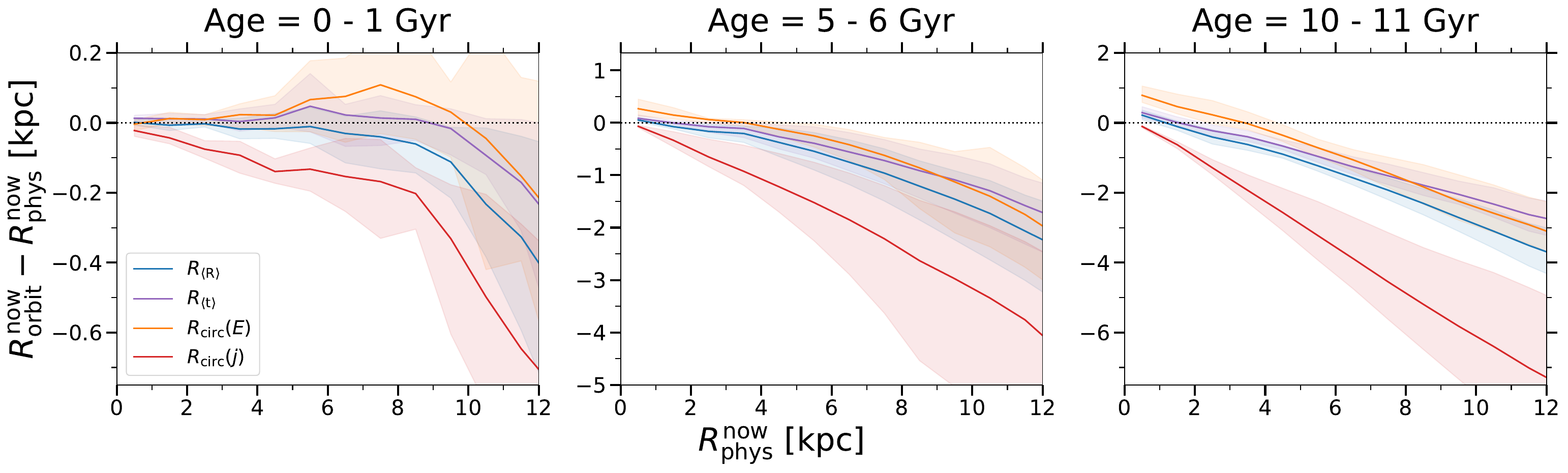}
\caption{
\textbf{Comparing the 4 metrics of the orbital radius of stars today against their instantaneous physical radius.}
Lines show the mean and shaded regions show the 68th percentile scatter across the 11 galaxies (8 for $\RE$ and $\Rj$).
Panels show different age stellar populations, corresponding (from left to right) to the late-disk, early-disk, and pre-disk eras \citep{McCluskey2024}.
The different orbital radii are most similar at smaller $\Rphys$ and for younger stars, while they are consistently smaller than $\Rphys$ at large $\Rphys$ and at older ages.
In general, $\RR$, $\Rt$, and $\RE$ yield similar results, while $\Rj$ is systematically smaller.
}
\label{fig:radius_metric}
\end{figure*}

We measure stellar orbits in a cylindrical coordinate system.
First, we calculate the center of each galaxy at each snapshot using an iterative zoom-in method on all star particles that end up within each MW-mass galaxy at $z = 0$ \citep[see][]{Wetzel2023}.
We then define the coordinate system with respect to the center at each snapshot, using the moment-of-inertia tensor of the youngest 25\% of stars that encompass 90\% of the stellar mass within 10 kpc of the galaxy to define the disk plane.
We use young stars as a compromise between using all stars and using gas, given that the morphology of old stars is generally not disky, and the morphology of gas can change rapidly during episodes of strong feedback.

Because stars are not on perfectly circular orbits, the definition of an `orbital radius', $\Rorbit$, is ambiguous, and various works use different metrics \citep[for example][]{Halle2015, ElBadry2018a, Lian2022, Okalidis2022}.
To comprehensively explore the effects on radial redistribution, we investigate $5$ metrics of $\Rorbit$:
\begin{itemize}
\item $\Rphys$: instantaneous physical radius
\item $\RR$: spatially-averaged orbital radius
\item $\Rt$: time-averaged orbital radius
\item $\RE$: radius of a circular orbit with the same energy
\item $\Rj$: radius of a circular orbit with the same angular momentum:
\end{itemize}

\textbf{$\Rphys$ is simply the instantaneous physical radius of a star particle relative to the galaxy's center.}

We also measure the \textbf{spatially-averaged orbital radius}, $\RR$:
\begin{equation}
\RR = \frac{R_{\rm apo} + \Rperi}{2}
\end{equation}
where $R_{\rm apo}$ is the apocenter of the orbit and $\Rperi$ is the pericenter.
We calculate this by tracking the radius of each star particle at each snapshot back to its birth and averaging adjacent apocenters and pericenters.

We also measure the \textbf{time-averaged orbital radius}, $\Rt$:
\begin{equation}
\Rt = \frac{1}{t_{\rm 2} - t_{\rm 1}} \int_{t_{\rm 1}}^{t_{\rm 2}} R(t) \, {\rm d} t
\end{equation}
where $t_{\rm 1}$ and $t_{\rm 2}$ are the times of an adjacent apocenter and pericenter (or vice versa).

\begin{figure}
\centering
\includegraphics[width = \columnwidth]{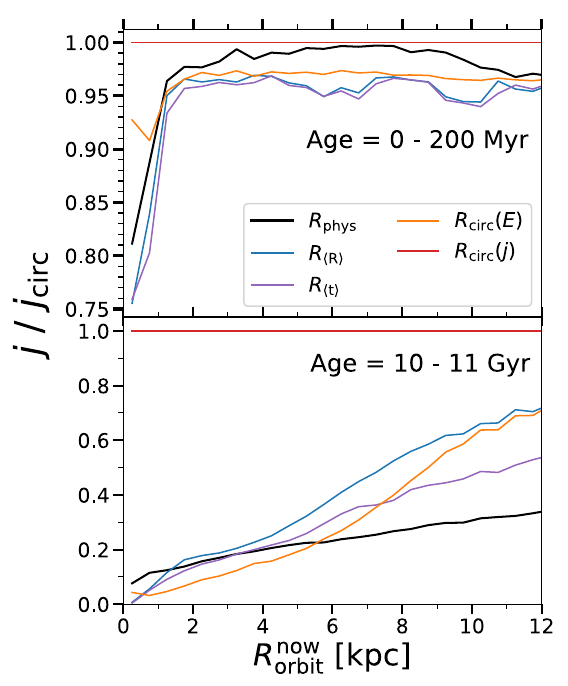}
\caption{
\textbf{Orbital angular momentum of stars today.}
\textbf{Top}: The median angular momentum of young stars (ages $< 200 \Myr$), relative to the angular momentum of a circular orbit at the same orbital radius, versus current orbital radius, $\Rorbitnow$.
By definition, $\Rj$ is unity.
Stars at $\Rorbitnow \lesssim 1.5 \kpc$ are on less circular, and more bulge/bar-like, orbits.
Beyond this radius, this ratio is approximately flat and $\gtrsim 0.95$; stars today form on nearly circular orbits.
\textbf{Bottom}: Same, but measuring orbits today for stars with ages $10 - 11 \Gyr$, which formed mostly during the pre-disk era.
This ratio increases with $R^{\rm now}_{\rm phys}$, indicating that old stars are on more circular orbits at larger radii, although the median is always $\lesssim 0.7$.
}
\label{fig:j_v_r}
\end{figure}

Following \citet{Abadi2003}, we also measure the \textbf{radius of a circular orbit with the same energy}, $\RE$:
\begin{equation}
\label{eqn:RE}
\Phi(R_{\rm circ}(E)) + \frac{1}{2} R_{\rm circ}(E) \frac{\,d\Phi(R)}{\,dR}\biggr\rvert_{R = R_{\rm circ}(E)} = E
\end{equation}
where $\Phi$ is the gravitational potential and $E$ is the sum of the kinetic and potential energy.
FIRE-2 simulations record the gravitational potential at the location of each particle at each snapshot.
We calculate the potential at a given $R$ (azimuthally-averaged) using the average potential of all star, gas, and dark-matter particles within $\pm 0.3 \kpc$ of the galactic disk midplane at each snapshot, in radial bins of width $250 \pc$ out to $R^\star_{90}$ \citep[see][]{Bellardini2022}.
We tested using the azimuthally-dependent potential and found variations in our inferred orbital radii of typically $\lesssim 5$\%.
For the potential beyond $R^\star_{90}$, we extrapolate assuming a constant slope of the potential with $R$.
We require the potential to increase monotonically with $R$.
In rare instances (typically at early times) when the potential decreases with $R$, we set its value equal to the value in the adjacent inner bin of $R$.
Using Equation~\ref{eqn:RE}, we calculate the energy of a circular orbit at different $R$.
We solve for $\RE$ using the numerical solution to the total energy of a circular orbit as a function of $R$ and the total energy of the star particle.

Our final metric is the \textbf{radius of a circular orbit with the same angular momentum}, $\Rj$:
\begin{equation}
\label{eqn:Rj}
R_{\rm circ}(j) = j^{2/3} \left[ \frac{\,d\Phi(R)}{\,dR} \biggr\rvert_{R = R_{\rm circ}(j)} \right] ^{-1/3}
\end{equation}
where henceforth we use $j$, where $j = \jphi = \vphi \Rphys$ is the specific angular momentum along the disk midplane (or equivalently, the azimuthal action).
Importantly, we self-consistently solve for the radial gradient of the potential following the same procedure as for $\RE$.
We use Equation~\ref{eqn:Rj} to calculate $j$ of a circular orbit at a given $\Rphys$, and we use this numerical solution to find $\Rj$ for a given $j$.
Many works start with the relation $j = v_{\rm circ}(r) R$ and use this to compute $\Rj$ given $j$, but doing this generally requires measuring $v_{\rm circ}(r) = \sqrt{G M_{\rm tot}(< r) / r}$ in spherical apertures, not just near the disk midplane, so our use of Equation~\ref{eqn:Rj} based on the gradient of the potential along cylindrical $R$ of the disk is more accurate.
$\Rj$ is the same as the `guiding center radius' used in the literature \citep[for example][]{SchoenrichBinney2009a, Frankel2020}.

Each of these metrics has advantages and disadvantages.
$\Rphys$ is straightforward to measure, and it most directly translates to the selection function of, for example, a magnitude-limited observational survey of the MW.
However, for a non-circular orbit, $\Rphys$ varies throughout the orbital phase, such that a star would experience some radial redistribution along a fixed orbit simply from its phase.
As we show below, this additional scatter from the orbital phase is important only for young stars, because older stars have experienced sufficient radial redistribution that the contribution from orbital phase is minor.

The other four metrics are, in principle, independent of orbital phase (assuming no perturbations).
$\RE$, and $\Rj$ are instantaneous measurements, so we can measure them immediately after a star particle forms (typically $\approx 10 \Myr$ after, given our snapshot time spacing) and today, thus providing full information about perturbations across a star particle's history.
However, these two metrics rely on instantaneous measurements of the (azimuthally average) gravitational potential, which in reality changes over the timescale of an orbit.
Furthermore, observationally $\RE$ and $\Rj$ are difficult to measure for stars, given uncertainties in the underlying gravitational potential.

By contrast, $\RR$ and $\Rt$ are independent of any approximations/assumptions about the potential or changes in orbital properties, such as angular momentum or total energy.
However, these two metrics require tracking/averaging a star particle across at least half of an orbit.
Thus, when we present these orbital radii for star particles `at birth' or `today', these obfuscate any initial and final changes to an orbit that occurred on a half-orbit timescale.
Furthermore, one cannot measure these two radii directly in observations.

We explore these 5 metrics of $\Rorbit$, to understand how much they affect inferences of radial redistribution, both in terms of the $\Rorbit$ used to measure radial redistribution and the $\Rorbit$ used to select stars today (or at birth).
As we will show, most of our results are broadly independent of the choice of metric for $\Rorbit$, \textit{as long as one selects stars and measures radial redistribution self-consistently.}

\begin{figure}
\centering
\includegraphics[width = \columnwidth]{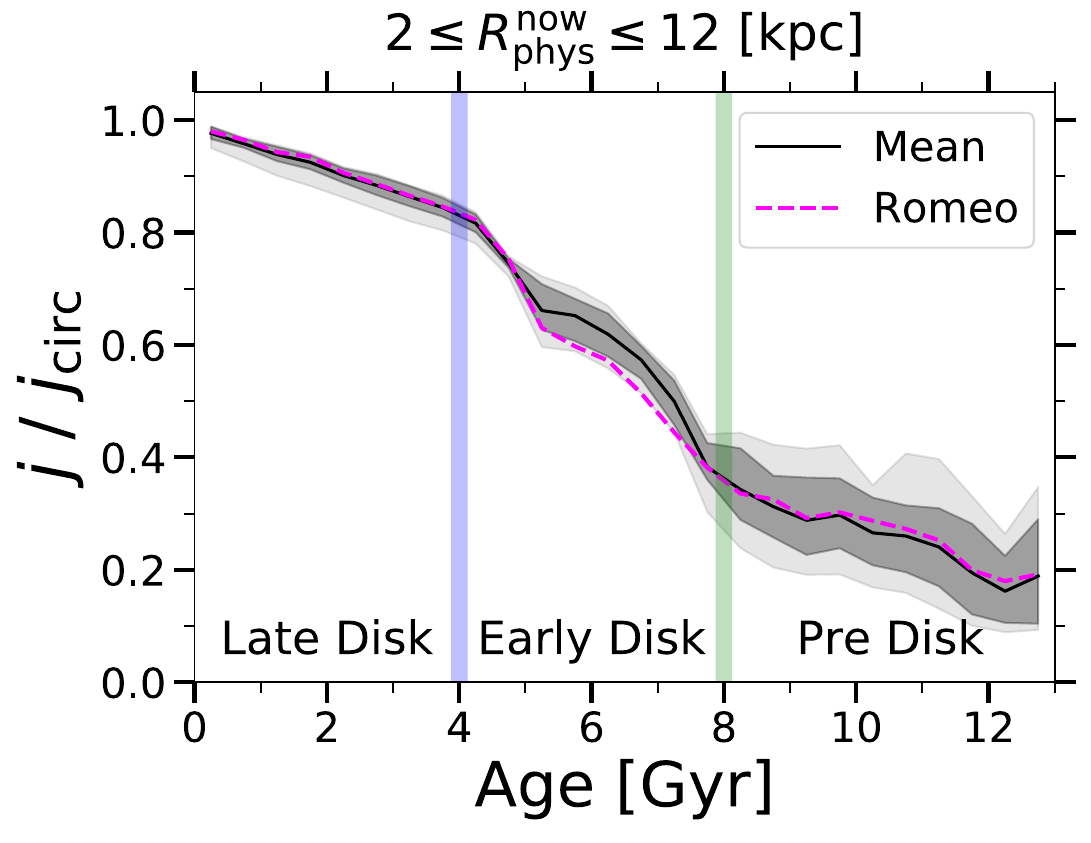}
\caption{
\textbf{Evolution of specific angular momentum, $j$.}
The median $j$ of stars relative to that of a circular orbit at the same $\Rorbitnow$ today, versus age, for all stars at $R^{\rm now}_{\rm phys} = 2 - 12 \kpc$ (excluding the bulge region), averaged across radial bins of width $1 \kpc$.
The black line shows the mean, while shaded regions show the 68th percentile and full scatter, across our 11 galaxies.
The dashed line shows Romeo, whose disk started to form the earliest, $\approx 11 \Gyr$ ago, but which follows a similar trend as the median.
The vertical bands separate the pre-disk (age $\gtrsim 8 \Gyr$), early-disk (age $\approx 4 - 8 \Gyr$), and late-disk (age $\lesssim 4 \Gyr$) eras.
Stars born in the pre-disk era are generally on highly eccentric orbits today.
The ratio increases rapidly in the early-disk era, and it flattens somewhat near unity during the late-disk era, when most stars were born on near-circular orbits.
}
\label{fig:j_v_age}
\end{figure}

\subsection{Measuring changes to orbits}

We measure each star particle's change in orbital radius, $\dRorbit$, specific angular momentum, $\Delta j$, and azimuthal velocity, $\Delta \vphi$, from its birth until today.
Unless otherwise specified, we then measure the median change of property $X$ across a given population as
\begin{equation}
\Delta X = \text{median}(X^\text{now} -  X^\text{birth})
\end{equation}
and we measure the half-width of the 68\% scatter in $\Delta X$ (about the median change) as
\begin{equation}
\sigma(\Delta X) = \frac{1}{2} \left(\text{percent}(\Delta X, 84) - \text{percent}(\Delta X, 16) \right)
\end{equation}
which equals the standard deviation for a normal distribution.

As we described in the Introduction, many analyses focus only on the \textit{scatter} in changes to orbital properties, such as $\sigma(\dRorbit)$, under the expectation that the median (net) change for a population is negligible \citep{SellwoodBinney2002, Frankel2018, Zhang2025bimodal}.
However, we investigate changes to both the median and the scatter about this, to highlight regimes when this common assumption does not hold.

We emphasize that we compute the \textit{median} (not mean) change in these quantities across all star particles within a given galaxy.
Stars can gain an arbitrary amount of $j$ but can lose only a finite amount (neglecting scattering to retrograde orbits).
Thus, scattering can lead to an asymmetric change, with a smaller number of stars each gaining a larger $j$, from a larger number of stars each losing a smaller $j$.

The formation radius of a star born `ex situ', in another galaxy that then merged into the host galaxy, is ill-defined.
Therefore, we include only stars that formed `in situ'.
Following \citet{Bellardini2022}, we define in-situ stars as those born within $30 \kpc$ comoving of the center of the main (most massive) progenitor of the MW-mass galaxy.
As Table~1 of \citet{Bellardini2022} showed, the ex situ fraction of stars within this radius today is small: $2-8\%$ across our sample.

We measure these quantities for each galaxy and show most results averaged across our 11 galaxies.
One simulation, Romulus \& Remus, does not have the gravitational potential stored in its snapshots, so we cannot measure $\Rj$ and $\RE$ accurately for those two galaxies.
Therefore, we show results for $\Rj$ and $\RE$ averaged over 9 galaxies instead of 11.

\section{Results}
\label{sec:results}

\subsection{Comparing metrics of orbital radius}
\label{sec:radius_compare}

We first compare our metrics of orbital radius, $\Rorbit$.
Figure~\ref{fig:radius_metric} shows the difference relative to the instantaneous radius, $\Rorbitnow - \Rphysnow$, versus $\Rphysnow$, for the four other metrics of $\Rorbit$ and for stars of different ages.
Stars on perfectly circular orbits would show no difference, while stars on more eccentric orbits would show larger differences.
For all metrics, the youngest stars agree best with $\Rphysnow$, because these stars are on the most circular orbits, especially at $\Rphysnow \lesssim 10 \kpc$.
The difference becomes more negative at larger $\Rphysnow$, where stars are more likely to be near their orbital apocenter, such that $\Rphysnow$ is maximal.
By contrast, $\Rorbitnow$ can be larger than $\Rphysnow$ at small $\Rphysnow$, where stars are preferentially closer to pericenter.

These trends persist at all stellar ages, though the discrepancies generally increase with age, reflecting the different eras of disk formation \citep{McCluskey2024}.
The oldest stars formed in the pre-disk era with the least circular orbits, while younger stars formed with increasingly coherent rotation.

These trends are generally similar across the 4 metrics for $\Rorbit$.
However, $\Rj$ is systematically the smallest at all ages.
This is because orbits generally have non-zero eccentricity, but $\Rj$ is sensitive \textit{only} to $j$, not to any non-circular motion.
In fact, $\Rj$ tends to be similar (within a factor of $\approx 2$) to the pericenter distance, $\Rperi$, because $j$ completely determines $\Rj$ and $j$ largely determines $\Rperi$.
(Considering limiting cases: for a circular orbit, $\Rperi = \Rj$, and for a maximally eccentric Keplerian orbit, $\Rperi = \Rj / 2$.)
By contrast, $\RE$ tends to be the largest, because it depends only on the orbital energy and, unlike the others, is insensitive to $j$ or eccentricity.
$\Rt$ and $\Rphys$ tend to skew closer to apocenter, because an orbit spends more time near apocenter, and $\Rphys$ is essentially a random sampling across time of the phase of an orbit.
As a result, $\RR$, which is insensitive to time sampling, is generally smaller than $\Rt$.

As an additional comparison of the impact of the metric of $\Rorbit$, in Appendix~\ref{sec:abundance_profile} we examine the radial profiles of metallicity for mono-age stellar populations today.
In summary, these population profiles depend only weakly on the choice of $\Rorbit$.
While the different metrics for $\Rorbit$ somewhat change the normalization of the metallicity at a given $\Rorbit$, they have a negligible effect on the radial gradient (the change with $\Rorbit$) for mono-age stars.

\begin{figure*}
\centering
\includegraphics[width = \linewidth]{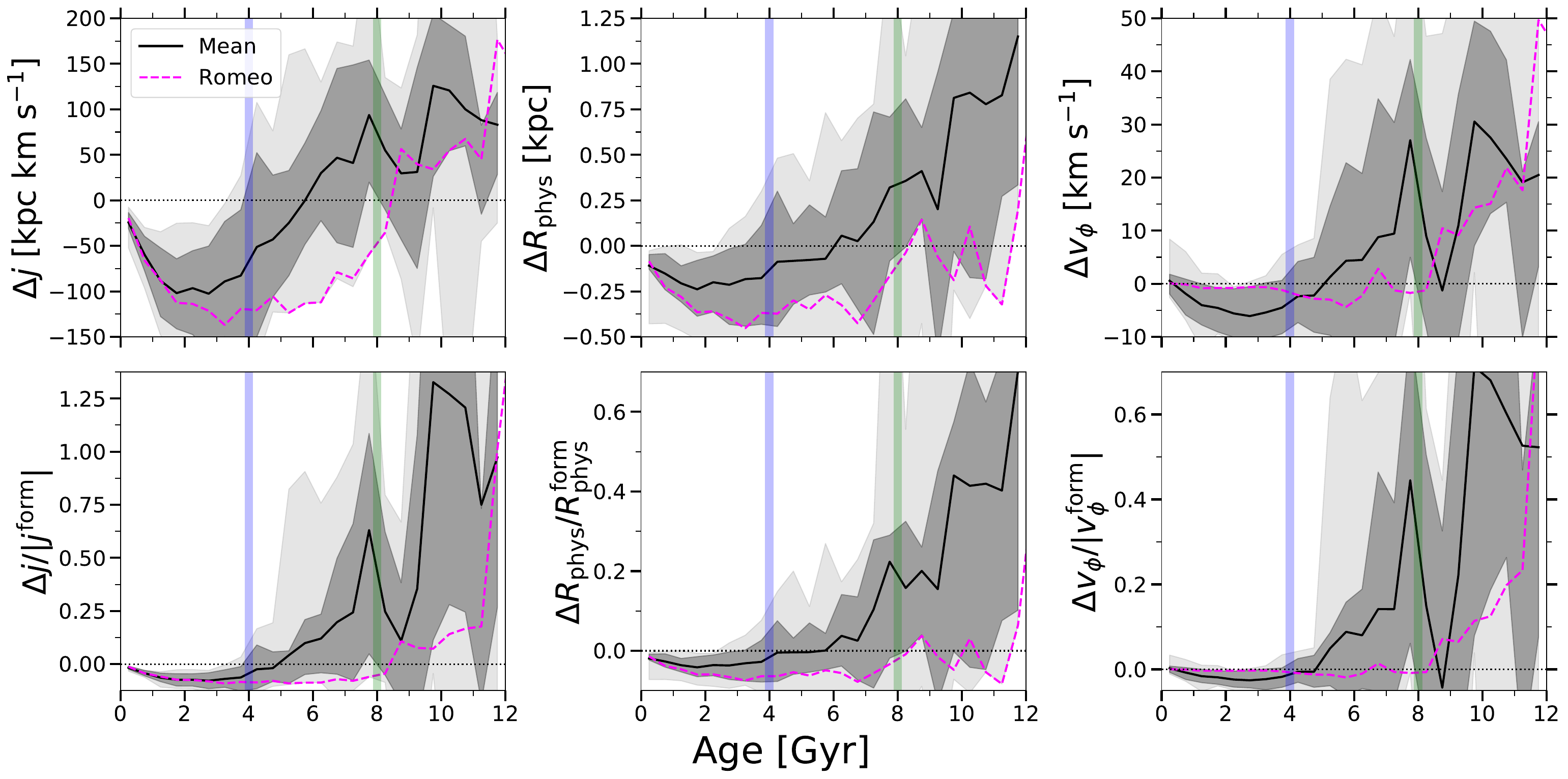}
\caption{
\textbf{Dynamical changes to stars from birth to today across the galaxy.}
The change in specific angular momentum, $j$ (left), physical radius, $\Rphys$ (middle), and azimuthal velocity, $\vphi$ (right), for all stars at $\Rphysnow = 2 - 12 \kpc$ (excluding the bulge region).
Solid lines show the mean and the shaded regions show the 68th percentile and full scatter across our 11 galaxies.
The dashed line shows Romeo, which has the earliest-forming disk that started to form $\approx 11 \Gyr$ ago.
The vertical bands separate the pre-disk, early-disk, and late-disk eras.
\textbf{Top row}: The median (net) change.
Stars that formed during the pre-disk era generally gained $j$, while stars that formed during the late-disk era generally lost it.
Stars that gained (lost) $j$ typically did so through an increase (decrease) in \textit{both} $\Rphys$ and $\vphi$.
\textbf{Bottom row}: Same, but for the fractional change relative to the value \textit{at birth}, which shows more distinctive behavior in each era and diminishes in magnitude over time, being negative but near zero during the late-disk era.
Stars that formed in the pre-disk and early-disk eras generally `spun up' afterwards, while stars that formed in the late-disk era generally `spun down' afterwards.
Romeo shows weaker changes at old ages, given its early-forming disk, though it experiences a stronger reduction in $j$ and $\Rphys$ during the early- and late-disk eras.
}
\label{fig:jrv_v_age}
\end{figure*}

\begin{figure*}
\centering
\includegraphics[width = 0.9 \linewidth]{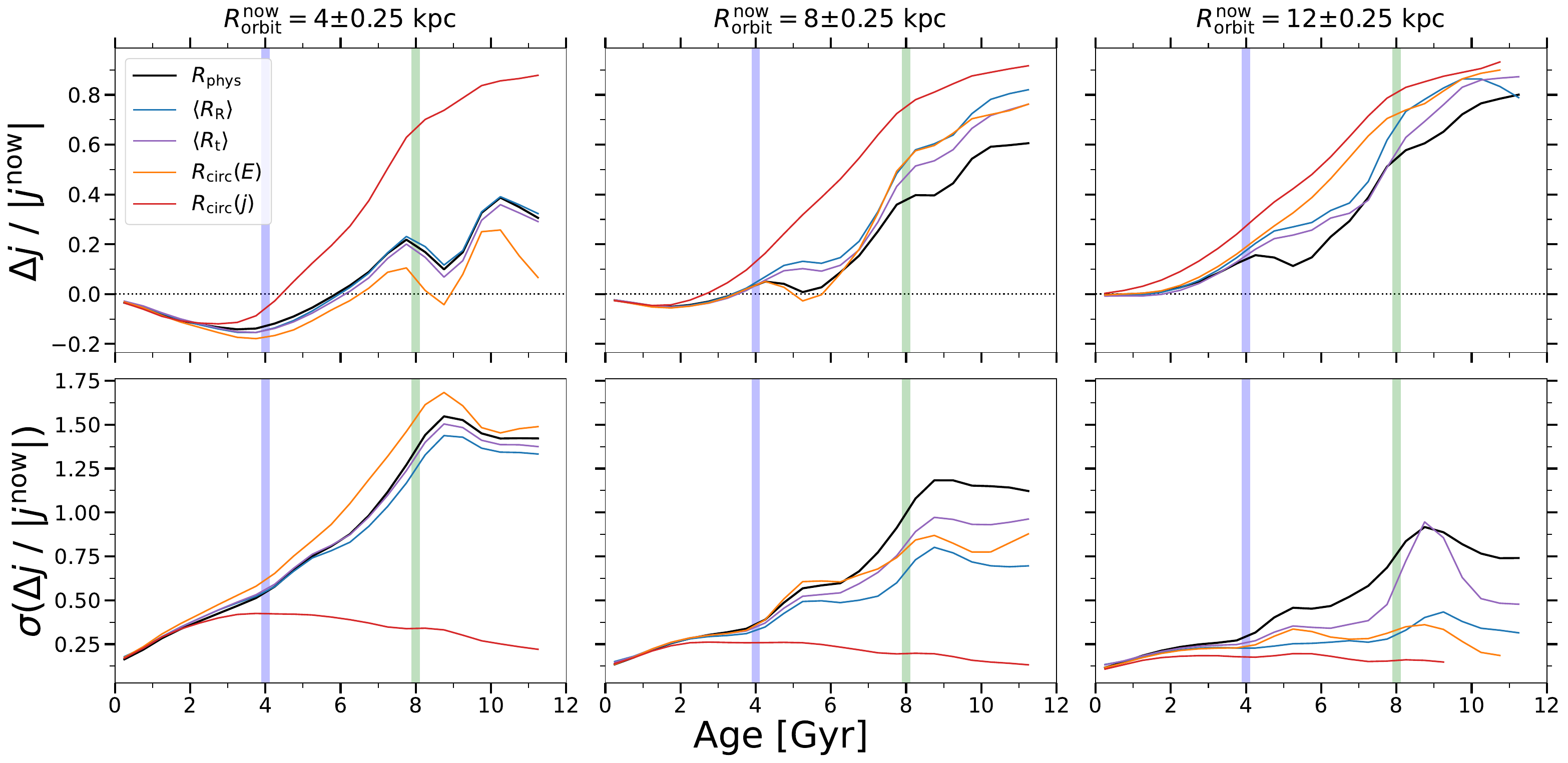}
\caption{
\textbf{Fractional changes to orbital angular momentum of stars, from birth to today, versus age}, relative to the value today, averaged over our galaxies.
Columns show stars at different $\Rorbitnow$ ($\pm 0.25 \kpc$), and each line shows a different metric for $\Rorbitnow$.
The vertical bands separate the pre-disk, early-disk, and late-disk eras.
\textbf{Top}: The median (net) fractional change.
Older stars and stars at larger $\Rorbitnow$ generally experienced the most significant increase.
By contrast, younger stars or stars at smaller $\Rorbitnow$ are more likely to have lost $j$ since birth.
If instead one selects stars at a given \textit{birth} radius, they typically lost $j$ at almost all ages (see Figure~\ref{fig:dj_v_age_rform}).
The age of transition, from $j$ increasing to $j$ decreasing since birth, is younger for stars at smaller $\Rorbitnow$.
\textbf{Bottom}: Same, but for the scatter (half width of the $68$th percentile) in the fractional change in $j$.
For all metrics of $\Rorbitnow$ except $\Rj$, this scatter increases with age.
For stars younger than $\approx 2 \Gyr$, all metrics of $\Rorbitnow$ yield similar results, but for older stars, measuring using $\Rj$ generally leads to more increase in the median $j$ (top) but also the smallest scatter (bottom).
}
\label{fig:dj_v_age_rnow}
\end{figure*}

\subsection{Evolution of angular momentum}
\label{sec:angular_momentum}

Next, we examine how each metric of $\Rorbit$ affects the profile of specific angular momentum, $j$, for stars today.
Figure~\ref{fig:j_v_r} (top) shows the ratio of $j$ to that of a circular orbit at the same $\Rorbitnow$, versus $\Rorbitnow$, for young stars (age $< 200 \Myr$).
By definition, this is unity for $\Rj$.
For all other metrics, $j / j_{\rm circ}$ is near unity at $\Rorbitnow \gtrsim 1.5 \kpc$, indicating that young stars in these FIRE-2 simulations generally formed on near-circular orbits \citep[see also][]{Yu2021, McCluskey2024}, at least beyond the inner (bulge) region.

Figure~\ref{fig:j_v_r} (bottom) shows the same but for old stars (ages $10 - 11 \Gyr$).
Now, for all metrics of $\Rorbit$ other than $\Rj$, $j / j_{\rm circ}$ increases with $\Rorbitnow$, likely because older stars at larger $\Rorbitnow$ were scattered to larger $j$ (as we explore below).
Furthermore, Figure~\ref{fig:j_v_r} shows the general trend that, at a given $\Rorbitnow$, $j / j_{\rm circ}$ decreases with age, because of the combined effects that younger stars formed on more circular orbits and older stars experienced more scattering and dynamical heating after birth \citep[see][]{McCluskey2024}.

To quantify this dependence on age directly, Figure~\ref{fig:j_v_age} shows the median $j / j_{\rm circ}$ across $\Rphysnow = 2 - 12 \kpc$ today as a function of stellar age.
We select $\Rphysnow > 2 \kpc$ to avoid the bulge region, beyond which $j / j_{\rm circ}$ is essentially constant for young stars, but we explore the impact of selection radii in Appendix~\ref{sec:radial_selection}.

The blue and green vertical bands demarcate the pre-disk (right), early-disk (middle), and late-disk (left) eras, from \citet{McCluskey2024}, as in Section~\ref{sec:transition_epochs}.
The shape of the evolution of $j / j_{\rm circ}$ correlates with the transitions, further supporting the meaning of these eras, even as measured for stars today.
Stars born in the pre-disk era have negligible $j$, stars born in the early-disk era have rapidly evolving $j$, and stars born in the late-disk era are on nearly circular orbits.

Figure~\ref{fig:j_v_age} separately shows Romeo, whose disk started to settle the earliest in our FIRE-2 simulations, $\approx 11 \Gyr$ ago \citep{McCluskey2024}.
Even though Romeo has the earliest-settling disk, its history as measured today is similar to the median across our galaxies.
This reinforces a key result from \citet{McCluskey2024}, that the present-day dynamical history, as one would measure from the orbits of stars today, does not directly reflect the dynamics of stars at birth, because orbits can change.

To quantify this, we next explore the change in $j$ of stars across the galaxy since their birth.
Figure~\ref{fig:jrv_v_age} (top) shows the median change in $j$ (left), as well as its components: $\Rphys$ (center), and $\vphi$ (right).
Again, we show the median of all stars across $\Rphysnow = 2 - 12 \kpc$ versus stellar age.

On average, stars born $\gtrsim 5.5 \Gyr$ ago experienced a net \textit{increase} in $j$, thus, they were torqued up.
This results from a net increase in \textit{both} $\Rphys$ and $\vphi$ since birth.
Older stars necessarily formed at systematically smaller $R$ (because the galaxy was smaller) and are more likely to have scattered to larger $R$.
The trend for $\vphi$ likely results from disk spin-up; during the pre-disk era, stars formed with little coherent rotational velocity, but as the disk settled, old stars scattered to larger $\vphi$ \citep{McCluskey2024}.
The increase in \textit{both} $\vphi$ and $\Rphys$ causes the net increase in $j$ since birth.

Stars $\lesssim 5.5 \Gyr$ old generally decreased in $j$ since birth, while stars older than this increased in $j$.
This transition age roughly corresponds to the middle of the early-disk era, when the disk settled enough to become `thin' (see also Figure~\ref{fig:j_v_r}), so perturbations generally reduced the relatively high $j$ of stars at birth.
The middle and right panels show that these changes in $j$ correspond to similar changes in \textit{both} $\Rphys$ and $\vphi$.

Interpreting absolute changes to these quantities can be difficult, given that typical values vary substantially across a galaxy's history, as its disk forms.
Thus, Figure~\ref{fig:jrv_v_age} (bottom row) shows the \textit{fractional} change with respect to the absolute value at birth.
Now, the trends with age correlate more clearly with the eras of disk formation.
Stars born in the pre-disk era generally experienced a large (order unity) fractional increase in $j$, $\Rphys$, and $\vphi$, because they formed with such small initial values that even small perturbations can lead to large fractional increases.
During the early-disk era, as the disk started to settle, the fractional changes were more modest.
Finally, in the late-disk era, stars formed on nearly circular orbits and experienced weak reductions in these quantities.

Figure~\ref{fig:jrv_v_age} separately shows Romeo: its transition from increasing to decreasing $j$, $\Rphys$, and $\vphi$, occurred at earlier ages, because its disk formed unusually early among our sample.
Furthermore, as the bottom panels show, its older stars experienced less fractional increase in $j$ since their birth, largely because its early disk onset caused its older stars to form with higher $j$ at birth.
Romeo's weaker fractional increase in $j$ since birth helps explain why its present-day history of $j / j_{\rm circ}$ in Figure~\ref{fig:j_v_age} is similar to the average galaxy in our sample, despite its earlier disk onset.

Figure~\ref{fig:jrv_v_age} (right) shows a sharp dip to $ \Delta \vphi \approx 0$ at approximately $9 \Gyr$, which also manifests itself in $j$ (left).
This likely reflects the dramatic fluctuations during the pre-disk era, when the galaxies did not have a well-defined and long-lived disk about which to compute these quantities in well-defined cylindrical coordinates.

Figure~\ref{fig:dj_v_age_rnow} (top) is similar to Figure~\ref{fig:jrv_v_age} (left bottom), except it shows the fractional change in $j$ versus age for stars currently at $\Rorbitnow = 4$, $8$, and $12 \kpc$ ($\pm 250 \pc$).
Here, we normalize the fractional change relative to $j$ today, rather than at birth.
Our motivation is to inform studies of stars in the MW and nearby galaxies, and to be consistent with now selecting stars at a given $\Rorbitnow$.
Thus, a value of $1$ means that all of the present-day $j$ of a star was gained after birth.

At all $\Rorbitnow$, older stars typically experienced the largest fractional change (increase) to $j$, because they formed with the smallest $j$, and they had the most time to experience redistribution.
Indeed, stars born in the pre-disk era generally gained most of their present $j$ since birth.
Furthermore, for all metrics of $\Rorbit$ except $\Rj$, stars of a given age at larger $\Rorbit$ experienced a larger increase in $j$ since birth.
The transition between typically gaining and typically losing $j$ occurred at older ages at smaller $\Rorbit$.
This reflects the inside-out radial growth of these galaxies \citep{Graf2025b}: stars at large $\Rorbit$ today are there because they gained (or did not lose much) $j$, whereas stars at small $\Rorbit$ could have lost substantial $j$, regardless of when they formed.

Figure~\ref{fig:dj_v_age_rnow} (bottom) shows the half-width of the 68\% scatter (comparable to $1 \sigma$) about the median fractional change in $j$, which quantifies the diffusion of $j$ since birth.
At a given age, this fractional scatter decreases with $\Rorbitnow$, largely because $j^{\rm now}$ generally increases with $\Rorbitnow$, such that the scatter in $\Delta j$ depends only weakly on $j^{\rm form}$ (similar to the trend that we will show for $\dRorbit$ below).
At a given $\Rorbitnow$, this fractional scatter in $j$ generally increases with age.

Figure~\ref{fig:dj_v_age_rnow} shows broadly similar trends for all metrics of $\Rorbit$, especially for the youngest stars, consistent with Figure~\ref{fig:radius_metric}.
The exception is $\Rj$ for stars older than $\approx 3 \Gyr$, which show systematically larger fractional changes in $j$.
This is because $\Rj$ tends to be the smallest $\Rorbit$ (Figure~\ref{fig:radius_metric}), and stars at a given $\Rj$ have the maximum $j$ (among all of the metrics) at that $\Rorbit$, so stars at a given $\Rj$ reflect the trends at larger $\Rorbit$ for the other metrics.
Similarly, because $\Rj$ is uniquely sensitive to $j$, the scatter in $\Delta j$ (bottom row) is minimal and is nearly independent of age, because all stars at that $\Rj$ have the same $j_{\rm now}$ by definition.

In summary, when selecting stars today at a given $\Rorbitnow$, regardless of the metric, older stars tend to have gained $j$ while younger stars tend to have lost $j$ since birth.
These trends for $j$ emerge from similar changes in \textit{both} $\Rphys$ and $\vphi$.
All metrics of $\Rorbit$ show similar fractional changes to $j$, with similar scatter, across all ages and all radii, except for $\Rj$, which shows systematically larger gains in $j$, and with systematically smaller scatter given the direct mapping between $j$ and $\Rj$.

That said, these trends are based on selecting stars at a given $\Rorbitnow$, as one would measure in the MW and nearby galaxies today.
However, in Appendix~\ref{sec:birth_radius} we show trends selecting stars at a given $\Rorbitform$.
This leads to significant differences, namely: (1) stars at a given $\Rorbitform$ generally lost $j$ since birth, regardless of age, and (2) the trends for $\Rj$ are similar to the other metrics of $\Rorbit$.

\subsection{Radial redistribution}
\label{sec:radial_redistribution}

\begin{figure*}
\centering
\includegraphics[width = \linewidth]{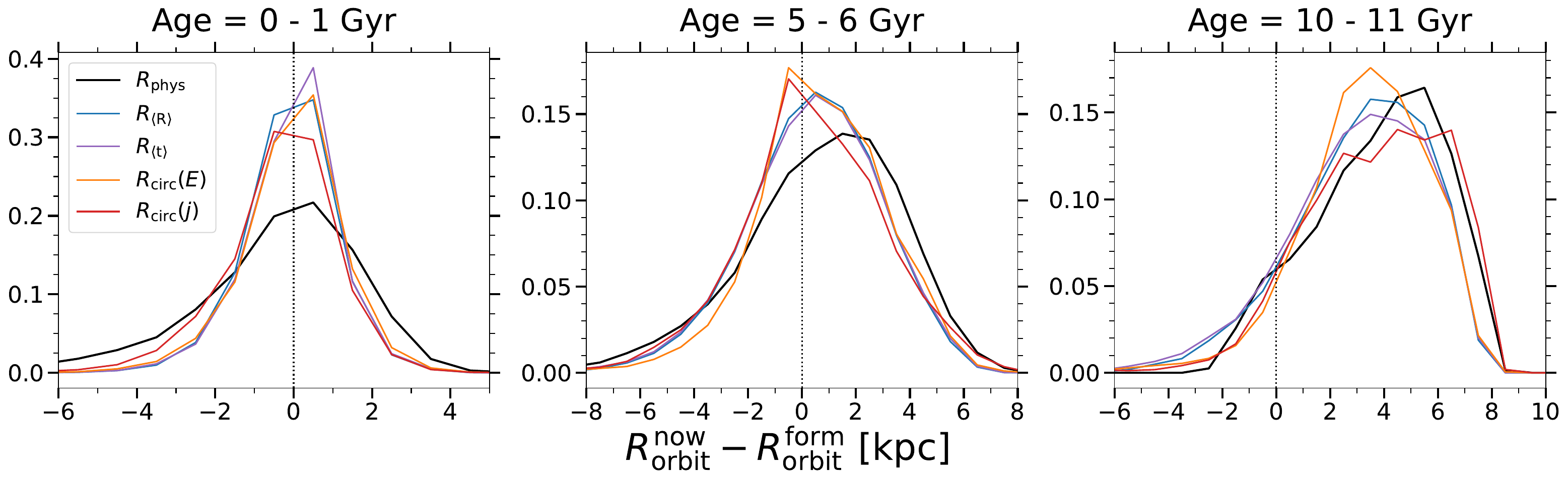}
\caption{
\textbf{Normalized distribution $\dRorbit$ of stellar orbits, from birth to today}, for stars at $\Rorbitnow = 8 \pm 0.25 \kpc$ today, averaged across our galaxies.
Because we normalize all distributions to unity, the scale of the y-axis is arbitrary.
Panels show stars of different ages today, which formed during the late-disk (left), early-disk (middle), and pre-disk (right) eras.
Each line shows a metric for $\Rorbit$ (at both birth and today): these yield similar results, except that using $\Rphys$ leads to a larger scatter for the youngest stars.
The median $\dRorbit \approx 0$ for young stars, but older stars experienced more redistribution to larger radii.
The width of the distribution is smaller for younger stars and increases to intermediate ages but is essentially unchanged for older stars.
}
\label{fig:dr_distr}
\end{figure*}

We next examine the changes to the orbital radius of stars from birth to today, $\dRorbit = \Rorbitnow - \Rorbitform$.
We first examine the full distribution of $\dRorbit$ for stars currently near the Solar annulus.
Figure~\ref{fig:dr_distr} shows the normalized distribution of $\dRorbit$ for stars at $\Rorbitnow = 8 \pm 0.25 \kpc$, at 3 ages, corresponding to stars that formed in the late-disk, early-disk, and pre-disk eras.
We show the 5 ways of measuring $\Rorbit$, self-consistently using the same metric to select stars today and to measure radial redistribution.

Figure~\ref{fig:dr_distr} (left) shows young stars today (ages $< 1 \Gyr)$.
The distributions are similar for all metrics of $\Rorbit$: symmetric and peaking near $\approx 0 \kpc$, with no net inward/outward redistribution.
Different metrics show similar scatter ($\approx 2.6 \kpc$), except the scatter in $\Rphys$ is $\approx 2 \times$ larger, because young stars have not experienced much change to their orbit, but $\Rphys$ has additional scatter from sampling phases of a non-circular orbit.

Figure~\ref{fig:dr_distr} (middle and right) shows radial redistribution for older stars, which experienced increasingly net outward redistribution toward larger $\Rorbit$.
This follows from the combined effects of the declining density profile of the galaxies (such that more stars scattered to a given $\Rorbitnow$ from smaller $\Rorbitform$ than from larger $\Rorbitform$), together with the significant dynamical heating of orbits for old stars in these simulations \citep[see][]{ElBadry2016}.
The distribution remains nearly symmetric at $\approx 5 \Gyr$, but it is negatively skewed for the oldest stars, for the reasons above.
The width of the distribution is larger than for young stars (left), but importantly, the width (amount of diffusion) is essentially unchanged from $\approx 5$ to $\approx 10 \Gyr$, which we discuss more below.

\subsection{Radial redistribution versus radius}
\label{sec:dr_vs_r}

\begin{figure*}
\centering
\includegraphics[width = \linewidth]{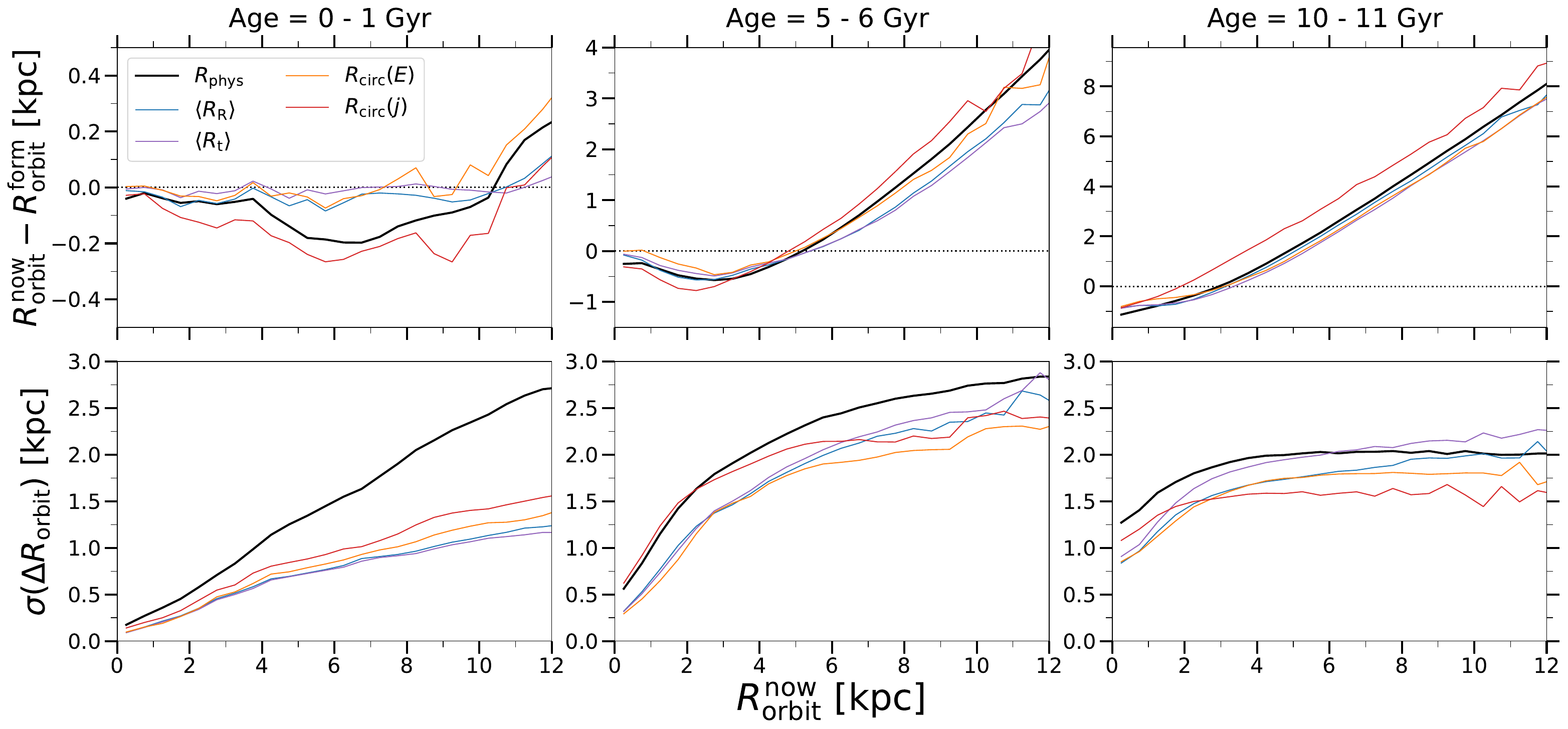}
\caption{
\textbf{Radial redistribution of stellar orbits, from birth to today, versus current orbital radius, $\Rorbitnow$}, averaged across our galaxies.
Panels show stars of different ages today, which formed during the late-disk (left), early-disk (middle), and pre-disk (right) eras.
Each line shows a different metric for $\Rorbit$ (at both birth and today).
\textbf{Top}: Median (net) change in $\Rorbit$.
Young stars typically moved slightly inward ($\lesssim 100 - 200 \pc$), while for older stars, those at larger $\Rorbitnow$ experienced more outward redistribution.
This is because older stars formed at systematically smaller $\Rorbit$ (the galaxy progenitor was more compact), so those at larger $\Rorbitnow$ necessarily redistributed outward.
\textbf{Bottom}: Same, but for the scatter (half width of the 68th percentile) of $\dRorbit$.
At all ages, the scatter initially increases with $\Rorbitnow$ but tends to saturate, especially for older stars, again given that they systematically redistributed outward from an initially more compact galaxy.
When self-consistently using the same metric of $\Rorbit$ to measure radial redistribution and orbital radius today, the different metrics yield similar results (unlike in Figure~\ref{fig:dr_v_age_rphys}), though $\sigma$ is usually largest for $\Rphys$.
}
\label{fig:dr_v_rnow}
\end{figure*}

We next examine trends of radial redistribution versus current orbital radius.
Figure~\ref{fig:dr_v_rnow} shows the median $\dRorbit$ and $\sigma(\dRorbit)$ of young, intermediate, and old stars (corresponding to those born in the late-disk, early-disk, and pre-disk eras) versus $\Rorbitnow$.
When measured self-consistently like this, the different metrics of $\Rorbit$ broadly agree, as we discuss further in Section~\ref{sec:dr_vs_age} and Appendix~\ref{sec:rphys}.
This is in part because, \textit{if one measures $\dRorbit$ and $\Rorbitnow$ self-consistently, with the same metric for $\Rorbit$, as we do here, then at a given $\Rorbitnow$, the only source of variation is $\Rorbitform$}.

At all ages, stars at smaller $\Rorbitnow$ typically moved inward since birth, while stars at larger $\Rorbitnow$ generally redistributed outward since birth.
The orbital radius at which stars transition from net inward to net outward redistribution is smaller with increasing age.
This reflects the inside-out radial growth of these galaxies \citep{Graf2025b}: older stars necessarily have smaller $\Rorbitform$, so the only way for them to have large $\Rorbitnow$ is via outward redistribution.
By contrast, younger stars had a larger range of $\Rorbitform$, so at a given $\Rorbitnow$, they are more likely to have moved inward than older stars.

For stars younger than $1 \Gyr$, for all metrics except $\Rphys$ and $\Rj$, $\dRorbit \approx 0 \kpc$.
$\Delta \Rphys$ and $\Delta \Rj$ show slight inward redistribution $\lesssim 0.2 \kpc$.
$\Rphys$ is uniquely sensitive to the phase of the orbits, which is the dominant effect of `redistribution' for young stars.
The mild decline in $\Rj$ since birth reflects the net loss of $j$ as in Section~\ref{sec:angular_momentum}, because stars were perturbed from their near-circular orbits at birth.

At older ages, $\dRorbit$ generally increases with $\Rorbit$, but the shape/slope changes with age.
Intermediate-age stars, formed in the early-disk era while the disk was settling, show a complex relation: $\dRorbit$ decreases with $\Rorbitnow$ in the inner galaxy and increases with $\Rorbitnow$ at $\gtrsim 3 \kpc$.
For the oldest stars, formed in the pre-disk era, the relationship is nearly linear: these stars all had small $\Rorbitform$, so $\dRorbit \approx \Rorbit$.

Figure~\ref{fig:dr_v_rnow} (bottom) shows $\sigma(\dRorbit)$, the scatter about the median, versus $\Rorbitnow$.
This is the most common method of quantifying radial redistribution: it reflects the combined inward and outward diffusion of stars about their birth radius, under an often implicit assumption that the median/net $\dRorbit = 0$.
Again, $\sigma(\dRorbit)$ is generally consistent across the metrics of $\Rorbit$, though again, it is systematically larger for $\Rphys$ for young stars because of the additional scatter from sampling the phase of an orbit.

At a given age, $\sigma(\dRorbit)$ generally increases with $\Rorbitnow$, though again, the shape/slope depends on age.
Given the inside-out radial growth of these galaxies, younger stars had a larger range of $\Rorbitform$, but they also had less time to redistribute.
For the youngest stars, this causes $\sigma(\dRorbit)$ to increase strongly with $\Rorbitnow$.
For the oldest stars, $\sigma(\dRorbit)$ is nearly flat with $\Rorbitnow$, because they all formed across a small range of $\Rorbitform$.
For intermediate-age stars, $\sigma(\dRorbit)$ increases out to $\approx 6 \kpc$, beyond which it is relatively flat.

Figure~\ref{fig:dr_v_rnow} shows results when selecting stars via $\Rorbitnow$, motivated by analyses of the MW and nearby galaxies.
Appendix~\ref{sec:birth_radius} presents similar results, but selecting stars via $\Rorbitform$.
These trends are generally opposite to those in Figure~\ref{fig:dr_v_rnow} (top), similar to the trends for $\Delta j$ in Section~\ref{sec:angular_momentum}: younger stars have $\dRorbit \approx 0$ or positive, and older stars generally have negative $\dRorbit$.
Again, the reason for this difference is that stellar density declines with $R$: most stars with a given $\Rorbitform$ moved inward, but when selecting stars by $\Rorbitnow$, more stars moved outward to that $\Rorbitnow$ rather than inward.
That said, Appendix~\ref{sec:birth_radius} shows that the trends of $\sigma(\dRorbit)$ are similar for selecting stars via $\Rorbitnow$ versus $\Rorbitform$.

\subsection{Radial redistribution versus age}
\label{sec:dr_vs_age}

\begin{figure*}
\centering
\includegraphics[width = \linewidth]{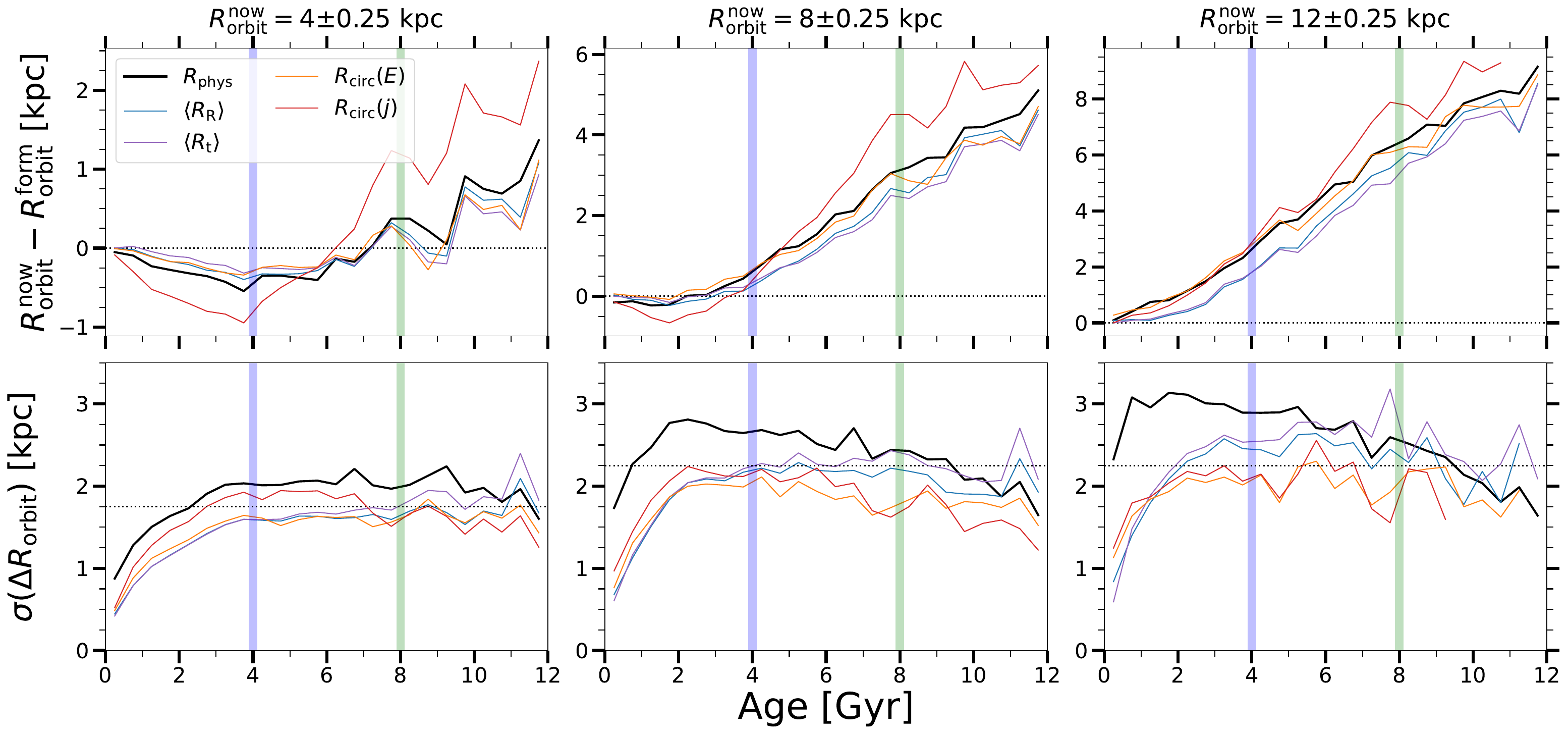}
\caption{
\textbf{Radial redistribution of stellar orbits, from birth to today, versus stellar age}, averaged across our galaxies.
Panels show stars at various $\Rorbitnow$.
The vertical bands separate the pre-disk, early-disk, and late-disk eras.
\textbf{Top}: The median (net) change in $\Rorbit$.
The youngest stars redistributed slightly inward, while older stars redistributed increasingly outward with age; this transition occurred earlier for stars at smaller $\Rorbitnow$.
If instead one selects stars at a given \textit{birth} radius, they typically decreased in $\Rorbit$ at almost all ages (see Figure~\ref{fig:dj_v_age_rform}).
\textbf{Bottom}: Same, but for the scatter (half width of the 68th percentile), $\sigma(\dRorbit)$.
Horizontal dotted lines show the average across all ages.
$\sigma(\dRorbit)$ initially increases with age up to $\approx 2 \Gyr$, beyond which \textit{the scatter in radial redistribution is approximately constant with age}, and in fact decreases slightly for older stars.
When self-consistently using the same metric of $\Rorbit$ to select stars today and measure redistribution, the different metrics yield similar results (unlike in Figure~\ref{fig:dr_v_age_rphys}), though $\sigma$ is systematically larger when using $\Rphys$ for younger stars.
}
\label{fig:dr_v_age_rnow}
\end{figure*}

Figure~\ref{fig:dr_v_age_rnow} shows trends in radial redistribution versus stellar age, for stars at various $\Rorbitnow$, and for various metrics of $\Rorbit$.
This figure highlights our most important results, as we articulate below.
Again, because we measure $\dRorbit$ and select stars via $\Rorbitnow$ self-consistently for each metric, at a given $\Rorbitnow$, the only source of variation is $\Rorbitform$.

Considering both $\dRorbit$ (top) and $\sigma(\dRorbit)$ (bottom), the different metrics of $\Rorbit$ lead to similar results.
Thus, \textit{the precise definition of $\Rorbit$ does not significantly affect the amount of radial redistribution, if one measures $\dRorbit$ and $\Rorbitnow$ self-consistently}.
In Appendix~\ref{sec:rphys}, we show that using the different metrics for $\dRorbit$ while selecting stars only via $\Rphysnow$, analogous to a simple observational selection, leads to much less similarity among the metrics for $\Rorbit$.

Figure~\ref{fig:dr_v_age_rnow} (top) shows median (net) $\dRorbit$ versus stellar age today, at various $\Rorbitnow$ ($\pm 0.25 \kpc$).
Generally, $\dRorbit$ is positive and increases with age.
However, for the youngest stars $\dRorbit$ is negative and decreases with age until $\approx 4 \Gyr$ ago at $\Rorbitnow = 4 \kpc$ (left) and until $\approx 2 \Gyr$ at $8 \kpc$ (middle).
This decrease in $\Rorbit$ for young stars follows from the decrease in $j$ in Section~\ref{sec:angular_momentum}, because young stars scattered from near-circular orbits at birth.
Again, the differing dependence on age at different $\Rorbitnow$, and the general increase in $\dRorbit$ for older stars, reflects the inside-out radial growth of these galaxies.
Older stars formed when the galaxy was smaller, so the only way for them to end up at large $\Rorbitnow$ is via outward redistribution, including via breathing modes induced by feedback from bursty star formation at early times \citep[see][]{ElBadry2016}.
At small $\Rorbitnow$, young stars (formed in the late-disk era) could have much larger $\Rorbitform$ and have moved inward, while old stars (formed in the pre-disk era) are more likely to have formed at even smaller $\Rorbitform$ and moved outward.

Furthermore, because stellar density declines with $R$, redistribution of stars from a given $\Rorbitform$ generally causes more stars at a given $\Rorbitnow$ to have come from smaller (rather than larger) $\Rorbitform$, leading to net positive redistribution at a given $\Rorbitnow$.
Indeed, as Appendix~\ref{sec:birth_radius} shows, stars born at a given $\Rorbitform$ tend to have moved inward, not outward, opposite to the general trends in Figure~\ref{fig:dr_v_age_rnow} (top), at least for $\Rorbitform \gtrsim 4 \kpc$.

Figure~\ref{fig:dr_v_age_rnow} (bottom) shows $\sigma(\dRorbit)$, the scatter of $\dRorbit$ about the median.
$\sigma(\dRorbit)$ is the most common metric of radial redistribution (or `radial migration'), quantifying the diffusion of stellar orbits about their birth $R$.
Young stars, born during the late-disk era, experienced the least redistribution, and $\sigma(\dRorbit)$ increases with age up to $\approx 2 \Gyr$.
However, \textit{at ages $\gtrsim 3 \Gyr$, $\sigma(\dRorbit)$ is relatively constant with age at all $\Rorbitnow$.}
Appendix~\ref{sec:birth_radius} shows that this flat dependence on age also holds when selecting stars via $\Rorbitform$, so it is not just a byproduct of selecting stars at $\Rorbitnow$.

Overall, Figure~\ref{fig:dr_v_age_rnow} highlights a key complementarity in the imprint of radial redistribution on $\dRorbit$ versus $\sigma(\dRorbit)$, according to the dynamical state of the disk when the stars were born.
For young stars, born during the late-disk era, $\sigma(\dRorbit)$ increases quickly with age, while $\dRorbit$ remains about constant with age.
This reflects the commonly expected behavior in a thin-disk regime of stellar orbits diffusing about their birth radii \citep[for example][]{SellwoodBinney2002}.
However, for stars born during the early-disk and pre-disk eras, $\gtrsim 3-4 \Gyr$ ago, $\sigma(\dRorbit)$ is nearly independent of age, while $\dRorbit$ increases monotonically with age.
Thus, \textit{for stars at $\Rorbitnow \gtrsim 8 \kpc$ born in the pre-disk era (before a long-lived disk formed) and in the earliest part of the early-disk era (when the disk was thick and dynamically hot), the primary mode of radial redistribution, from birth to today, is a net change in the median $\Rorbit$, rather than diffusion about the median, $\sigma(\dRorbit)$.}

\subsection{Comparison to previous works}
\label{sec:comparison}

\begin{figure}
\centering
\includegraphics[width = \columnwidth]{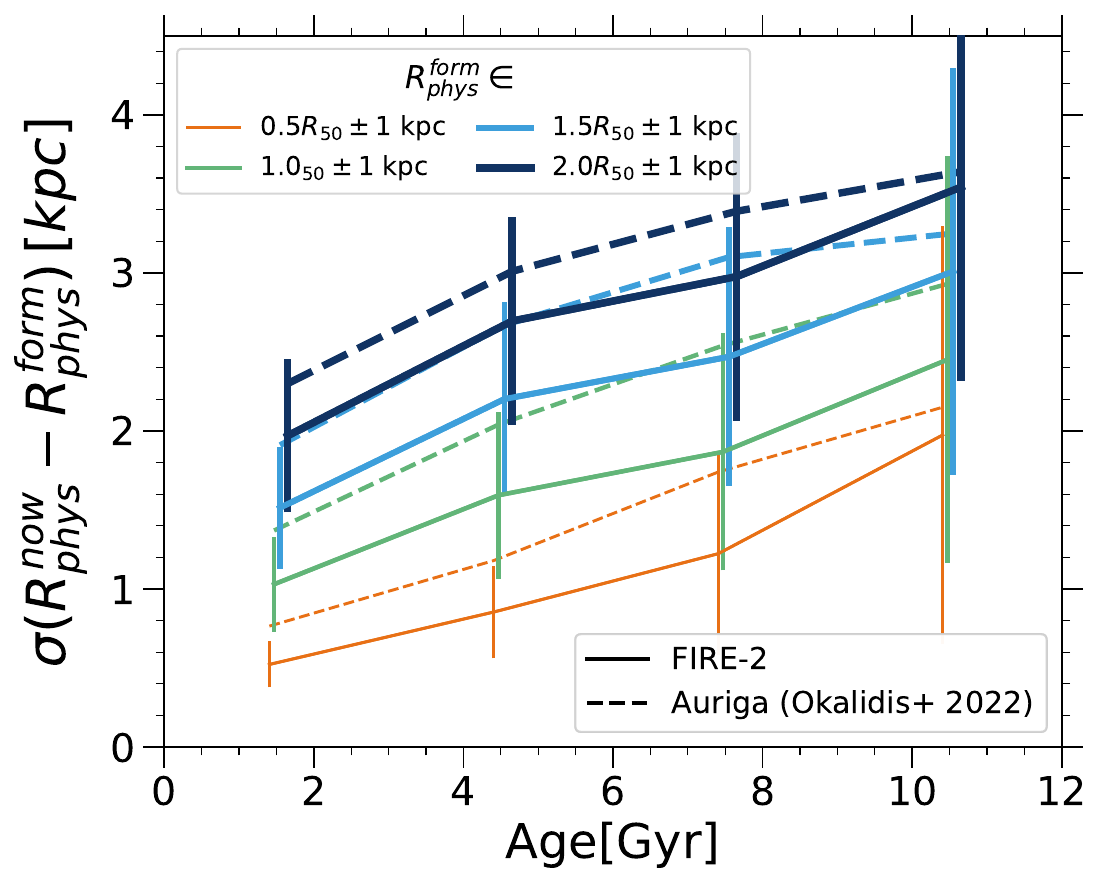}
\caption{
\textbf{Comparing FIRE-2 against the Auriga simulations.}
Radial redistribution, $\sigma(\dRorbit)$ (half width of the 68th percentile), versus stellar age, in the FIRE-2 simulations (solid lines) and the Auriga simulations (dashed lines) in \citet{Okalidis2022}.
Error bars show the $1 \sigma$ scatter across the simulation suite.
We select stars as in \citet{Okalidis2022}, matching their selection in $\Rphysform$ via multiples of $R^\star_{50}$ \citep[see][for additional details]{Bellardini2021}.
In FIRE-2, the typical $R^\star_{50} \approx 3 \kpc$ today, so these lines span $\Rphysform \approx 1.5 - 6 \kpc$.
This selection via $\Rphysform$ at smaller radii leads to a more monotonic increase with age than in Figure~\ref{fig:dr_v_age_rnow}.
Despite differences in modeling the ISM, star formation, and feedback, radial redistribution in FIRE-2 is broadly consistent (within the galaxy-to-galaxy scatter) with the Auriga simulations, though overall there is slightly less redistribution in FIRE-2.
}
\label{fig:literature_compare_auriga}
\end{figure}

We next place our results in the context of previous theoretical work.
We first compare with the results that \citet{Okalidis2022} presented from the Auriga cosmological zoom-in simulations \citep{Grand2017}.
This comparison provides a useful benchmark for stellar radial redistribution in a set of simulations that use significantly different models for the ISM, star formation, and feedback.

Figure~\ref{fig:literature_compare_auriga} compares $\sigma(\Delta \Rphys)$ in FIRE-2 to that measured in \citet{Okalidis2022} (their Figure~7) from 17 Auriga MW-mass galaxies.
Here, we reproduce the methodology of \citet{Okalidis2022}, whose selection differs from ours in several respects.
First, and most crucially, we now select stars based on birth radius, $\Rphysform$.
(We explore the impact of selecting via $\Rorbitform$ in more detail in Appendix~\ref{sec:birth_radius}.)
Second, we now select stars at multiples of each galaxy's present-day $R^\star_{50}$, as opposed to selecting based on fixed $\Rorbit$ across different galaxies, and we now use a slightly smaller vertical range of $|Z_{\rm now}| < 2 \kpc$ (instead of $|Z_{\rm now}| < 3 \kpc$).
Third, \citet{Okalidis2022} sought to measure only stars in the dynamically cold, thin disk and exclude stars in the bulge and halo, so we now impose their cut on present-day orbital circularity, $\epsilon > 0.7$.
Finally, we also match their larger age binning of 3 Gyr.

We tested how each of these new selection criteria impacts our results, finding that selection via $\Rphysform$ has the largest impact (see also Appendix~\ref{sec:birth_radius}), leading to stronger dependence on age.
In other words, $\sigma(\Delta \Rphys)$ does not flatten as much with age.
The other selection effects lead to only slight quantitative changes.
The average $R^\star_{50}$ of our FIRE-2 galaxies is $\approx 3 \kpc$, so their chosen radii (in multiples of each galaxy's $R^\star_{50}$) for FIRE-2 probes $\Rphysform \approx 1.5 - 6 \kpc$, on average smaller than the radial ranges we explored previously, leading to both smaller $\sigma(\Delta \Rphys)$ (because it increases with $\Rorbit$) and a stronger increase with age.
Including only stars with $\epsilon > 0.7$ leads to only a slight increase (5-8\% on average) in $\sigma(\Delta \Rphys)$.

The key point of Figure~\ref{fig:literature_compare_auriga} is that, when measuring radial redistribution with the same selection, the results from the FIRE-2 and Auriga simulations are broadly similar.
Notably, $\sigma(\Delta \Rphys)$ agrees within the $1 \sigma$ galaxy-to-galaxy scatter at most ages and $\Rphysform$.
However, FIRE-2 shows systematically \textit{less} radial redistribution than Auriga.
One possible origin of this difference is the galaxies' sizes: despite having similar stellar masses, the average $R^\star_{50}$ of FIRE-2 galaxies is smaller, $\approx 3 \kpc$, compared to $\approx 4 \kpc$ in \citet{Okalidis2022}.
Because in both works $\sigma(\Delta \Rphys)$ increases with $\Rorbit$, this partially could explain a discrepancy in normalization.
Furthermore, differences in modeling the ISM, star formation, and stellar feedback in the simulations also likely plays a key role, such that at early times, FIRE-2 simulations experience a clumpier ISM, burstier star formation, and stronger fluctuations in the gravitational potential, while conversely, at later times, bars are stronger and more prominent in Auriga.

While Figure~\ref{fig:literature_compare_auriga} compares $\sigma(\Delta \Rphys)$, we also verified that the trends for $\Delta \Rphys$ are broadly similar between the two simulation suites, at least for younger stars.
\citet{Okalidis2022} found that in Auriga, stars currently at $R < R^\star_{50}$ generally experienced net inward redistribution (average $\Delta \Rphys \approx -1 \kpc$), while stars currently at $R > R^\star_{50}$ generally experienced net outward redistribution (average $\Delta \Rphys \approx 1 \kpc$), which generally agrees with FIRE-2 in Figures~\ref{fig:dr_v_rnow} and \ref{fig:dr_v_age_rnow}.

\begin{figure}
\centering
\includegraphics[width = \columnwidth]{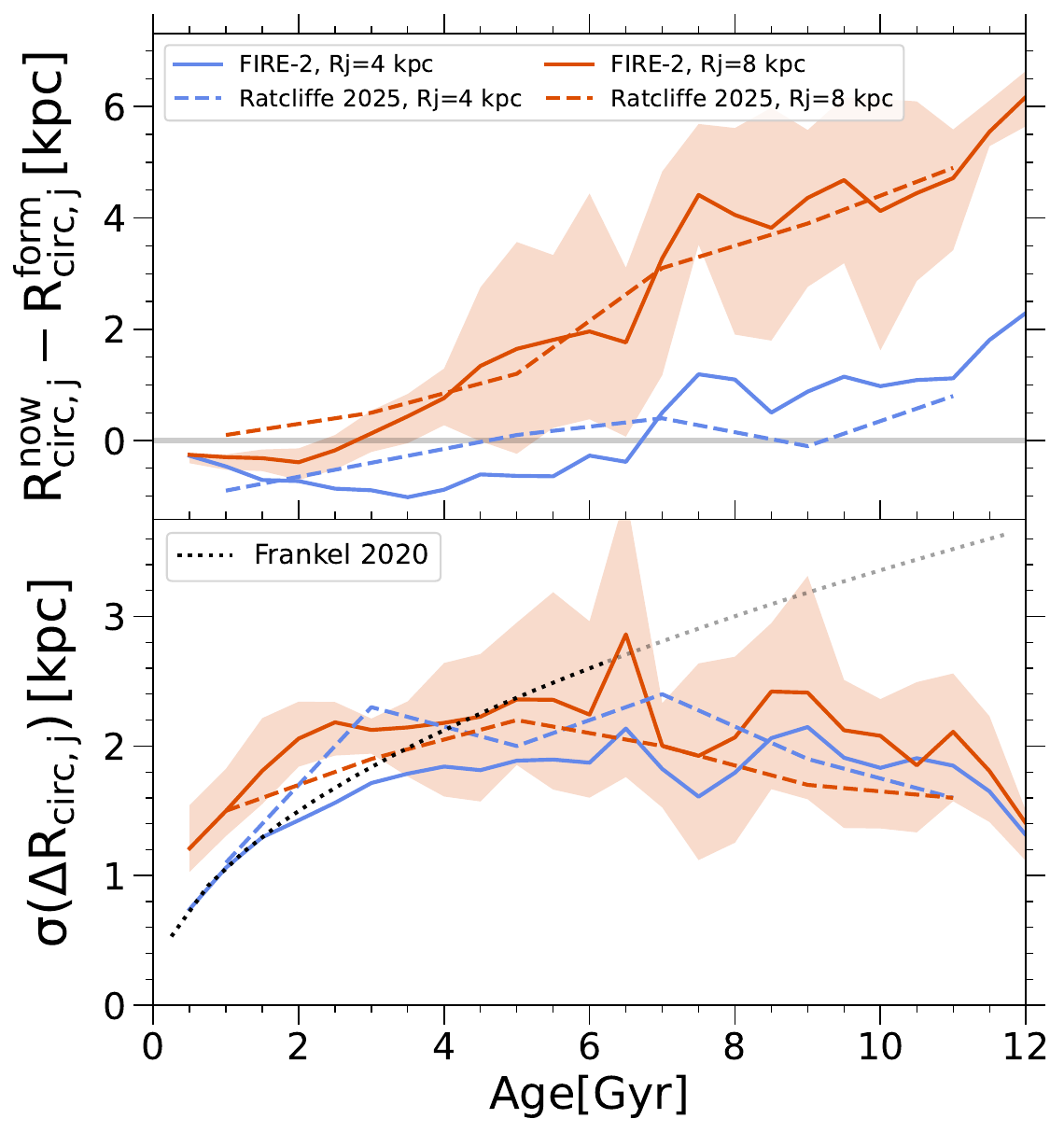}
\caption{
\textbf{Comparing radial redistribution in FIRE-2 with estimates for the MW}, for changes to `guiding-center radius', $\Rj$, versus stellar age today.
We select stars at $\Rj = 4$ and $8 \kpc$ ($\pm 1 \kpc$) today. Solid lines show the mean, and the shaded regions
show the 68th percentile (for $\Rj = 8 \kpc$) across our FIRE-2 galaxies.
\textbf{Top}: Net redistribution, in terms of median $\Delta \Rj$.
Dashed lines show the semi-empirical estimates for the MW from \citet{Ratcliffe2025a}, with agree well with our FIRE-2 results.
\textbf{Bottom}: Same, but for $\sigma(\Delta \Rj)$, the half width of the 68th percentile.
In addition to \citet{Ratcliffe2025a}, we also compare with the analytic model applied to the MW from \citet{Frankel2020}.
Again, FIRE-2 agrees well with \citet{Ratcliffe2025a}: both FIRE-2 and \citet{Ratcliffe2025a} show an increase with ages up to $3-4 \Gyr$ and a saturation (and possible decline) of $\sigma(\Delta \Rj) \lesssim 2 \kpc$ at older ages.
The increase with age in FIRE-2 is broadly consistent with \citet{Frankel2020} up to $3 - 4 \Gyr$, and plausible even up to $6 \Gyr$ (for $\Rj = 8 \kpc$), the age limit of the analysis in \citet{Frankel2020}.
However, both FIRE-2 and \citet{Ratcliffe2025a} are inconsistent with any implied monotonic increase from an \textit{extrapolation} of the fit from \citet{Frankel2020} beyond $6 \Gyr$, as the lighter grey dotted line shows.
\textit{Overall, radial redistribution in FIRE-2 agrees well with these observational estimates of the MW.}
}
\label{fig:literature_compare_mw}
\end{figure}

We next compare our results with empirically-derived inferences for the MW, specifically, the angular-momentum diffusion model of \citet{Frankel2020} and the metallicity-based reconstruction of stellar birth radii from \citet{Ratcliffe2025a}.
While measuring radial redistribution in simulations is relatively straightforward, any measurement in the MW must be inferred indirectly from the present-day structure in metallicity-dynamics of its stellar populations.
Such reconstructions necessarily rely on modeling assumptions about the past state of the MW.
Despite these uncertainties, recent efforts to `reverse engineer' the MW's dynamical history provide valuable semi-empirical constraints.

\citet{Frankel2020} developed a forward model to reconstruct the MW’s radial redistribution history from the present-day distribution of stars observed by Gaia and APOGEE.
In their framework, stars formed with a time- and radius-dependent metallicity that traces the evolving ISM, and subsequently redistributed through the disk via a diffusion process in guiding-center radius, $\Rj$.
The model integrates these prescriptions forward in time, generating the present-day joint distribution of stellar ages, metallicities, and $\Rj$, which they compare with the MW through Bayesian inference.
Importantly, \citet{Frankel2020} only considered the evolution on the `thin' (low-$\alpha$) disk, adopting a maximum age of $6 \Gyr$ for their model.

\citet{Ratcliffe2025a} used a semi-empirical method to infer stellar birth $R$ from age and metallicity, based on previous frameworks \citep{Minchev2018, Lu2022}.
Their model links the observed scatter in [Fe/H] across age bins to the underlying ISM metallicity gradient but incorporates a correction for the evolving width of the star-forming disk, calibrated on MW-mass galaxies from the TNG50 simulation.
\citet{Ratcliffe2025a} argued that this extension allows their method to recover the evolution of the MW's metallicity gradient and reconstruct the redistribution of stars across the disk.

Figure~\ref{fig:literature_compare_mw} compares the FIRE-2 simulations with the MW results of \citet{Frankel2020} and \citet{Ratcliffe2025a}.
The top panel shows the median change in `guiding-center radius', $\Rj$, versus stellar age today, comparing to just \citet{Ratcliffe2025a} (\citealt{Frankel2020} did not infer this quantity).
The bottom panel shows the corresponding half-width of the 68th-percentile, $\sigma(\Delta \Rj)$, versus stellar age today, comparing to both \citet{Frankel2020} and \citet{Ratcliffe2025a}.
We show two representative radial ranges from \citet{Ratcliffe2025a}, $4$ and $8 \kpc$, using a bin width $\pm 1 \kpc$.
\citet{Frankel2020} inferred only a single, radially-independent $\sigma(\Delta \Rj)$, so we show only one line for their model.
Furthermore, \citet{Frankel2020} presented an age-dependent analytic fit only at ages $< 6 \Gyr$.
So, we show their results at these measured ages in dark black and an extrapolation of their fit to older ages in lighter grey.

In Figure~\ref{fig:literature_compare_mw} (both panels), FIRE-2 agrees remarkably well with \citet{Ratcliffe2025a}.
Net $\Delta \Rj$ is minimal for young stars, but at $\Rj = 8 \kpc$ today it increases to $\Delta \Rj \approx 5 \kpc$ at ages $\gtrsim 10 \Gyr$.
Similarly, at $\Rj = 4 \kpc$ today, both works show that young stars experienced slight inward redistribution.
In the bottom panel, both \citet{Ratcliffe2025a} and FIRE-2 show that $\sigma(\Delta \Rj)$ increases with age but only up to $\approx 4 \Gyr$, and for older stars it flattens (and possibly decreases), saturating at $\lesssim 2 - 2.5 \kpc$.
Therefore, both FIRE-2 and \citet{Ratcliffe2025a} suggest that radial redistribution in terms of $\sigma(\Delta \Rj)$ is most effective early in a star’s life but does not continue to grow indefinitely with age.

\citet{Frankel2020} infer that $\sigma(\Delta \Rj)$ increases monotonically with age, $\tau$, as $\sigma(\Delta \Rj) = 2.6~{\rm kpc} \,(\tau / 6~{\rm Gyr})^{1/2}$.
A precise quantitative comparison with \citet{Frankel2020} is difficult, because their model is radially independent.
Nevertheless, the overall amplitudes of FIRE-2 and \citet{Frankel2020} (and \citealt{Ratcliffe2025a}) are broadly similar at ages $\lesssim 4 \Gyr$, where our two $R$ bins in FIRE-2 bracket \citet{Frankel2020}.
At $\Rj = 8 \kpc$ in FIRE-2, considering also the galaxy-to-galaxy scatter, the trends in FIRE-2 are consistent with \citet{Frankel2020} up to $6 \Gyr$, the age limit of their analysis.
Figure~\ref{fig:literature_compare_mw} also shows an extrapolation of the fit from \citet{Frankel2020} to older ages (in lighter grey), to emphasize that our results are not consistent with an extrapolated monotonic increase with age from their fit.
Therefore, one should \textit{not} assume that the strength of radial redistribution in terms of $\sigma(\Delta \Rj$) increases monotonically with age indefinitely.

\citet{Ratcliffe2025a}, \citet{Frankel2020}, and our FIRE-2 simulations all agree that $\sigma(\Delta \Rj) \approx 2 \kpc$ for stars $\approx 4 \Gyr$ old, suggesting a characteristic amplitude of radial diffusion in MW-like disks.
This is broadly consistent with the results of \citet{Lian2022}, who found $\sigma(\Delta \Rj) = 2.25$ and $1.85 \kpc$ at $\Rj = 4 \pm 1$ and $8 \pm 1 \kpc$ in the MW today for $3 \Gyr$ old stars.
In Section~\ref{sec:summary_and_discussion}, we further discuss these results in comparison to a variety of previous studies.

Finally, we acknowledge that the excellent agreement between FIRE-2 and \citet{Ratcliffe2025a} for the MW in Figure~\ref{fig:literature_compare_mw} may be surprising, for a few reasons.
First, our FIRE-2 galaxies were not selected to be exact MW analogs, and they can differ from the MW in some key respects.
FIRE-2 galaxies typically form disks later than the MW \citep{Yu2023, McCluskey2024}, generally exhibit hotter kinematics in both stars and gas \citep{McCluskey2025}, lack strong bars \citep{Ansar2025}, and have higher late-time star formation rates \citep{Gandhi2022}.
Thus, the relative impact and timing of various mechanisms that drive radial redistribution may differ in the FIRE-2 galaxies versus the MW.
For example, in the Pre-Disk Era, feedback-driven fluctuations in the gravitational potential drove coherent net outward redistribution in these FIRE-2 galaxies \citet{ElBadry2016}, while some works argue that the formation and slow-down of the MW's bar drove substantial outward redistribution $\approx 8 \Gyr$ ago \citet{Haywood2024}. 

Second, \citet{Ratcliffe2025a} inferred the degree of radial redistribution in MW via its effect on metallicity radial gradients, and in their framework, the MW's metallicity radial gradient for stars at birth was steeper in the past than today.
By contrast, the metallicity radial gradients of stars at birth in these FIRE-2 simulations were shallower in the past than today \citep{Bellardini2021, Graf2025a}.
To be clear, these FIRE-2 galaxies do agree with the MW that the metallicity radial gradient is shallower for older stars as measured today \citep{Graf2025a}, but the interpretation of this trend is different.
In FIRE-2, this observed trend in the MW qualitatively reflects the evolution of the star-forming ISM, while \citet{Ratcliffe2025a} assumed that the increase in stellar radial redistribution with age flattened what was a steeper gradient in the star-forming ISM for the oldest stars, such that redistribution inverts the trend with age, as measured today.
Thus, it may be surprising that our results agree so well with the inference for the MW in \citet{Ratcliffe2025a} in Figure~\ref{fig:literature_compare_mw}, given that the results in \citet{Ratcliffe2025a} are based on their model for the evolution of the MW's metallicity radial gradient.

The resolution to this is unclear.
One possibility is an offset of competing effects.
If the MW has an unusually early-forming disk, it may have experienced stronger redistribution early on from the onset of structures like bars and spiral arms.
By comparison, for a later-forming disk, the dynamical processes that keep a disk thicker and more turbulent longer themselves may drive stronger dynamical heating and redistribution of stellar orbits.
In principle, these two scenarios should imprint themselves differently in the nature of radial redistribution, specifically in the amount of cold torquing versus heating, and we will explore this in Steel et al. (in prep.).
While more work is needed to understand in detail the relationship between radial redistribution and a disk's dynamical history, as a first step for FIRE-2, in the next Section, we explore possible correlations between the strength of radial redistribution and disk formation times in FIRE-2.

\subsection{Correlations between radial redistribution and galaxy properties}
\label{sec:correlation}

\begin{table*}
\caption{
correlations between galaxy properties and the amount of radial redistribution stars experienced
}
\centering
\begin{tabular}{l|rrrrr|rrrrr|rrrrr}
\toprule
\multicolumn{1}{c|}{\multirow{3}{*}{Radial metric}} & 
\multicolumn{5}{c|}{$t_{\rm lb}[(\vphi / \sigma_{v \rm ,3D})_{\rm form} > 1]$} &
\multicolumn{5}{c|}{$R^\star_{90}$} &
\multicolumn{5}{c}{$M^\star_{90}$} \\
\multicolumn{1}{c|}{} &
\multicolumn{2}{c}{$R = 7 - 9 \kpc$} &
\multicolumn{1}{c}{} &
\multicolumn{2}{c|}{$R = 2 - 12 \kpc$} &
\multicolumn{2}{c}{$R = 7 - 9 \kpc$} &
\multicolumn{1}{c}{} &
\multicolumn{2}{c|}{$R = 2 - 12 \kpc$} &
\multicolumn{2}{c}{$R = 7 - 9 \kpc$} &
\multicolumn{1}{c}{} &
\multicolumn{2}{c}{$R = 2 - 12 \kpc$} \\
\multicolumn{1}{c|}{} &
\multicolumn{1}{c}{$\rho$} &
\multicolumn{1}{c}{p-val} &
\multicolumn{1}{c}{} &
\multicolumn{1}{c}{$\rho$} &
\multicolumn{1}{c|}{p-val} &
\multicolumn{1}{c}{$\rho$} &
\multicolumn{1}{c}{p-val} &
\multicolumn{1}{c}{} &
\multicolumn{1}{c}{$\rho$} &
\multicolumn{1}{c|}{p-val} &
\multicolumn{1}{c}{$\rho$} &
\multicolumn{1}{c}{p-val} &
\multicolumn{1}{c}{} &
\multicolumn{1}{c}{$\rho$} &
\multicolumn{1}{c}{p-val} \\
\midrule
$\Delta \Rphys$ & \textbf{-0.69} & \textbf{0.02} & &         -0.53  &            0.09 
& \textbf{-0.71} & \textbf{0.02} & & \textbf{-0.76} &    \textbf{0.01} 
& \textbf{-0.61} & \textbf{0.05} & & \textbf{-0.67} &    \textbf{0.02} \\
$\Delta \RR$    & \textbf{-0.68} & \textbf{0.02} & &         -0.47  &            0.15 
& \textbf{-0.61} & \textbf{0.05} & & \textbf{-0.79} & \textbf{0.01} 
& \textbf{-0.86} & \textbf{0.02} & & \textbf{-0.63} &    \textbf{0.04} \\
$\Delta \Rt$    & \textbf{-0.77} & \textbf{0.01} & &         -0.41  &            0.22 
& \textbf{-0.65} & \textbf{0.03} & & \textbf{-0.75} &    \textbf{0.01}  
& \textbf{-0.76} & \textbf{0.01} & &         -0.56  &            0.08  \\
$\Delta \RE$    & \textbf{-0.71} & \textbf{0.03} & &         -0.20  &            0.60 
&         -0.59  &         0.09  & &         -0.65  &            0.06   
& \textbf{-0.80} & \textbf{0.01} & &         -0.37  &            0.33  \\
$\Delta \Rj$    &          0.01  &         0.98  & &         -0.63  &            0.07 
&         -0.26  &         0.50  & & \textbf{-0.79} &    \textbf{0.01} 
&          0.08  &         0.83  & & \textbf{-0.80} &    \textbf{0.01} \\
\midrule
$\sigma (\Delta \Rphys)$ &         0.11  &            0.75   & &         0.14  &         0.68 
& \textbf{0.82} & \textbf{0.01} & & \textbf{0.64} & \textbf{0.04} 
&         0.26  &            0.45   & & \textbf{0.68} & \textbf{0.02} \\
$\sigma (\Delta \RR)$    &         0.31  &            0.36   & &         0.18  &         0.61 
& \textbf{0.69} &    \textbf{0.02}  & &         0.55  &         0.08 
&         0.51  &            0.11   & & \textbf{0.71} & \textbf{0.02} \\
$\sigma (\Delta \Rt)$    &         0.15  &            0.66   & &         0.18  &         0.61 
&         0.59  &            0.06   & &         0.55  &         0.08 
&         0.55  &            0.08   & & \textbf{0.71} & \textbf{0.02} \\
$\sigma (\Delta \RE)$    &         0.42  &            0.26   & &         0.22  &         0.57 
& \textbf{0.68} &    \textbf{0.05}  & &         0.64  &         0.06 
&         0.23  &            0.55   & &         0.55  &         0.13 \\
$\sigma (\Delta \Rj)$    &         0.52  &            0.15   & &         0.56  &         0.12 
& \textbf{0.90} & \textbf{0.01} & & \textbf{0.79} & \textbf{0.01} 
&         0.48  &            0.19   & & \textbf{0.70} & \textbf{0.04} \\
\bottomrule
\end{tabular}
\tablecomments{
Correlation strength, $\rho$, and significance (p-value), between median (net) $\dRorbit$, or $\sigma(\dRorbit)$, for all stars against: the lookback time to disk onset, $t_{\rm lb}[(\vphi / \sigma_{v \rm,3D})_{\rm form} > 1]$; the radius encompassing 90\% of the stellar mass today, $R^\star_{90}$; the associated stellar mass today, $M^\star_{90}$.
Correlation strengths and p-values that are statistically significant (p-value $\leq 0.05$) are bold.
We select stars in the Solar annulus ($\Rorbitnow = 7 - 9 \kpc$) and stars across the disk ($\Rorbitnow = 2 - 12 \kpc$).
The first column lists the metric of orbital radius used.
Generally, we find statistically significant correlations (p-value $< 0.05$) with $R^\star_{90}$ and $M^\star_{90}$.
$\dRorbit$ correlates negatively with galaxy size and mass, while $\sigma(\dRorbit)$ correlates positively with them.
The age of disk onset (when a long-lived disk first formed) only meaningfully correlates with $\dRorbit$ for stars in the Solar annulus today.
In particular, $\sigma(\dRorbit)$, the most commonly used metric of radial redistribution, does not correlate with the age of disk onset.
}
\label{tab:correlation}
\end{table*}

\begin{figure}
\centering
\includegraphics[width = \linewidth]{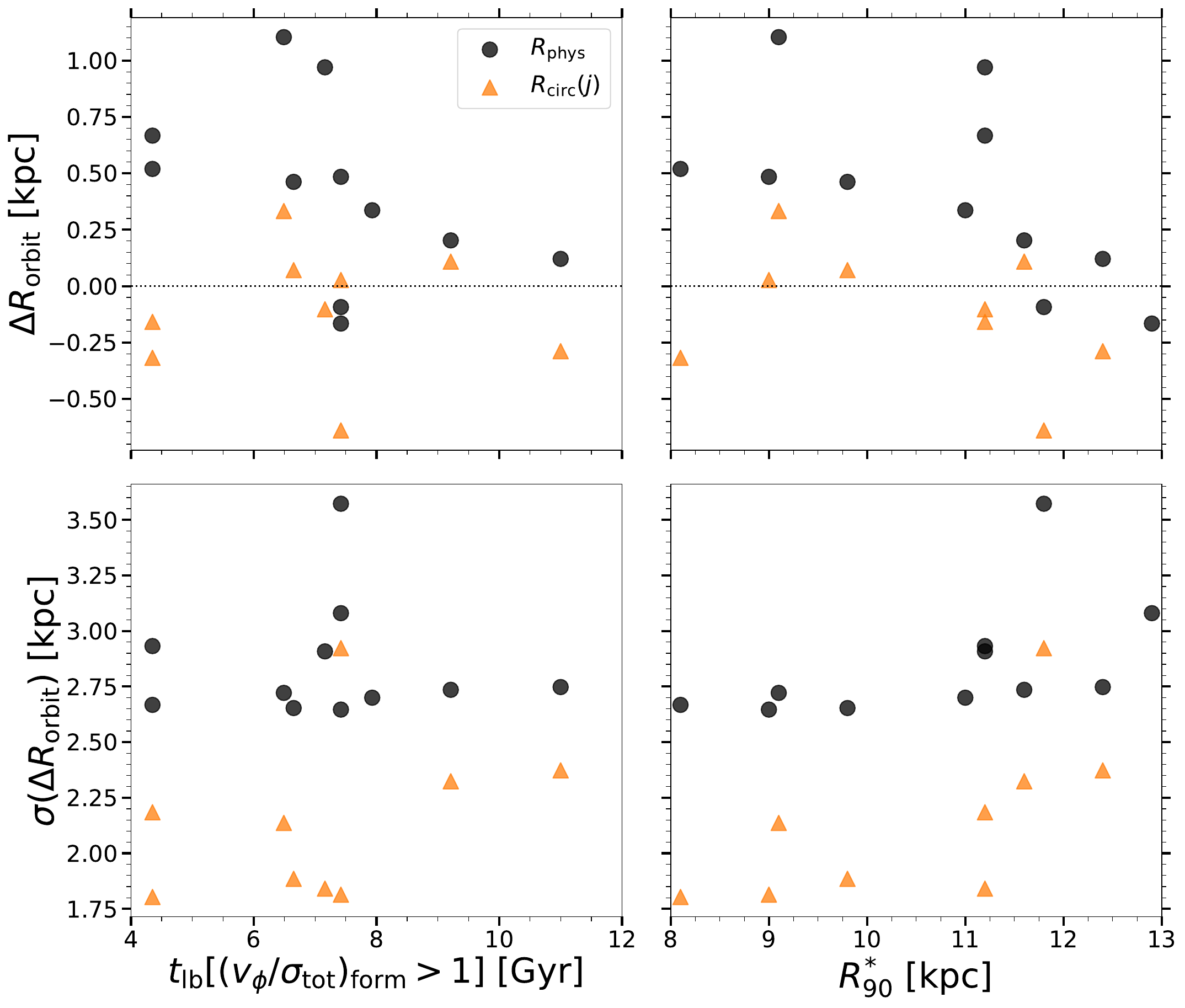}
\caption{
\textbf{Correlations of radial redistribution with galaxy properties today}, for all stars at $\Rorbitnow = 7 - 9 \kpc$.
We show correlations only with $\Rphys$ and $\Rj$, but Table~\ref{tab:correlation} shows correlations for all metrics of orbital radius; correlations with $\Rphys$ are similar to all other metrics except $\Rj$.
\textbf{Top}: Median (net) radial redistribution, $\dRorbit$, versus the lookback time of disk onset \citep[from][]{McCluskey2024}, $t_{\rm lb}[(\vphi / \sigma_{v \rm,3D})_{\rm form} > 1]$ (left), and the radius that encompasses 90\% of the stellar mass, $R^\star_{90}$.
For all metrics of $\Rorbit$ except $\Rj$, disks that formed later (left) or that are smaller today (right) experienced more outward radial redistribution.
\textbf{Bottom}: Similar, but for the scatter (half width of the 68th percentile), $\sigma(\dRorbit)$.
In general, \textit{radial redistribution in terms of $\sigma(\dRorbit)$ does not correlate significantly with when the disk started to form}.
However, $\sigma(\dRorbit)$ does correlate with $R^\star_{90}$, such that larger galaxies experienced more radial redistribution.
}
\label{fig:correlation}
\end{figure}

Finally, we test correlations between the amount of radial redistribution experienced and various galaxy properties today.
These correlations can inform the mechanisms driving radial redistribution, although we defer a detailed investigation of the dynamical processes to future work.

Table~\ref{tab:correlation} shows the Spearman rank correlations of the amount of radial redistribution (both in terms of the median and the scatter), using each metric of $\Rorbit$, for all stars currently near the Solar neighborhood, $\Rorbitnow = 7 - 9 \kpc$, and for stars across the disk, $\Rorbitnow = 2 - 12 \kpc$.
We examine 3 galaxy properties: (1) the lookback time to disk onset, when each galaxy transitioned from the pre-disk to the early-disk era, from \citet{McCluskey2024} based on $t_{\rm lb}[(\vphi / \sigma_{v \rm,3D})_{\rm form} > 1]$, (2) the galaxy size today, $R^\star_{90}$ of all stars, and (3) the galaxy stellar mass today, $M^\star_{90}$, with the latter two from Table~1 of \citet{Bellardini2021}.

Figure~\ref{fig:correlation} shows a visual sample of the correlations against the time of disk onset (left) and the size of the galaxy (right), for $\Rphys$ and $\Rj$.
We do not show the other metrics of $\Rorbit$, because their correlations are generally similar to $\Rphys$, and we do not show the trend versus $M^\star_{90}$.

In general, for stars near the Solar neighborhood, $\dRorbit$ correlates significantly (p-value $< 0.05$) and negatively with the transition time from the pre-disk to early-disk era, for example, as Figure~\ref{fig:correlation} (top left panel) shows for $\Rphys$.
That is, galaxies that formed their disk later show stronger net outward $\dRorbit$.
The likely reason is the stellar-feedback-induced breathing modes undergone by these galaxies during their pre-disk, bursty-star-formation era \citep{ElBadry2016, Yu2021}, which heat the orbits of stars: the later the disk formed, the longer this occurred, leading to more outward $\dRorbit$.
The exception to this correlation is $\Rj$, likely because these breathing modes tend to heat the orbits of stars primarily radially rather than torque them.

However, none of the above correlations are strong when including all stars across the disk today, presumably because stars that end up at small $\Rorbitnow$ cancel out any effect.
Furthermore, and even more importantly, we find no significant correlation between $\sigma(\dRorbit)$ and the time of disk onset, for any metric of $\Rorbit$ or for either selection range of $\Rorbitnow$.
Thus, \textit{earlier-forming disks do not experience more radial redistribution in terms of $\sigma(\dRorbit)$.}
This may be surprising, given that many mechanisms thought to drive $\sigma(\dRorbit)$ are specific to disk morphology, like scattering off of spiral arms and bars \citep[for example][]{Sellwood2014}.
However, this result is consistent with Figure~\ref{fig:dr_v_age_rnow} (bottom), that $\sigma(\dRorbit)$ depends on age only over the last few Gyr.

In terms of correlations with galaxy size and mass, Table~\ref{tab:correlation} (middle) and Figure~\ref{fig:correlation} (right) show that $\dRorbit$ correlates \textit{negatively} with galaxy size: smaller galaxies experienced more radial redistribution.
This could be because more compact galaxies have higher stellar densities, driving more scattering, though Table~\ref{tab:correlation} (right) shows broadly similar negative correlations for galaxy stellar mass.
Broadly speaking, these correlations hold both near the Solar neighborhood and across the disk.

$\sigma(\dRorbit)$ shows a more complicated relation with galaxy size and mass.
For stars in the Solar neighborhood, $\sigma(\dRorbit)$ correlates positively with galaxy radius: larger galaxies show more spread in radial redistribution.
This remains true, though to a lesser extent, for stars across the disk.
However, $\sigma(\dRorbit)$ experienced by stars near the Solar neighborhood does not correlate with galaxy stellar mass, while it does correlate strongly and positively with stars across the disk, such that more massive galaxies experienced more $\sigma(\dRorbit)$.

Furthermore, although we do not show it, we test correlations of the median $\dRorbit / R^\star_{90}$ and $\sigma(\dRorbit / R^\star_{90})$ against $R^\star_{90}$.
The correlations between $\dRorbit / R^\star_{90}$ and galaxy size are essentially the same as between the mean radial change and $R^\star_{90}$.
The correlations between $\sigma(\dRorbit / R^\star_{90})$ and galaxy size tend to be weaker than the correlations between $\sigma(\dRorbit)$ and galaxy size.

\section{Summary and Discussion}
\label{sec:summary_and_discussion}

\subsection{Summary}
\label{sec:summary}

We characterized radial redistribution, including changes to specific angular momentum and azimuthal velocity, of stars from birth to today, using 11 MW/M31-mass galaxies from the FIRE-2 suite of cosmological zoom-in simulations.
We comprehensively tested $5$ metrics of measuring galactocentric orbital radii:
\begin{itemize}
\item $\Rphys$, instantaneous physical radius
\item$\RR$, spatially-averaged radius (between apocenter and pericenter)
\item $\Rt$, time-averaged radius (between apocenter and pericenter)
\item $\Rj$ radius of a circular orbit with the same angular momentum
\item $\RE$, radius of a circular orbit with the same energy
\end{itemize}
We examined changes to both median properties, such as $\dRorbit$, and the scatter about this, such as $\sigma(\dRorbit)$, the latter being the most commonly used metric for radial redistribution (or `radial migration').
We also examined their dependence on current radius, formation radius, and stellar age.
Our key results are as follows. \\

\textbf{Comparing metrics of orbital radius, $R_{\rm orbit}$}:

\begin{itemize}
\item \textit{The instantaneous physical radius, $\Rphys$, is generally larger than other metrics of $R_{\rm orbit}$ today} (Figure~\ref{fig:radius_metric}).
$\Rphys$ is most similar to other metrics for young stars (typical differences $\lesssim 0.2 \kpc$), because they are on nearly circular orbits.
$\Rphys$ is as much as $3 - 7 \kpc$ larger than the other metrics (on average) for older stars.

\item \textit{If one uses the same definition of $\Rorbit$ to select stars today and measure radial redistribution, the results are largely insensitive to the metric of $\Rorbit$ used} (Figures~\ref{fig:dr_v_rnow} and \ref{fig:dr_v_age_rnow}).
By contrast, these metrics show stronger differences if selecting stars today based on $\Rphys$ (Figure~\ref{fig:dr_v_age_rphys}).
\textit{Therefore, one should use the same radius metric to select stars today and measure radial redistribution.}

\item \textit{Using different metrics of $\Rorbit$ leads to similar radial gradients in metallicity} (differences $\lesssim 0.005$ dex kpc$^{-1}$, Figure~\ref{fig:abundance_profile}).
Thus, the metric of $\Rorbit$ used does not significantly affect the metallicity radial gradient, but it can modestly shift the normalization of the metallicity profile at a given $\Rorbit$.
\end{itemize}

\textbf{Changes to orbital radius, $\Rorbit$, and specific angular momentum, $j$, since birth:}

\begin{itemize}
\item \textit{Changes in specific angular momentum, $j$, generally coincide with similar changes, in the same direction, to both orbital radius, $\Rorbit$, and azimuthal velocity, $\vphi$} (Figures~\ref{fig:jrv_v_age}, \ref{fig:jrv_v_age_radius}).
Therefore, we summarize results for changes to $j$ and $\Rorbit$ together.

\item \textit{Net changes to the median $\Rorbit$ and $j$ of a stellar population can be significant} (Figures~\ref{fig:jrv_v_age}, \ref{fig:dj_v_age_rnow}, \ref{fig:dr_v_rnow}, \ref{fig:dr_v_age_rnow}).
In fact, the primary mode of radial redistribution for stars born $\gtrsim 4 - 8 \Gyr$ ago (depending on radius today) is a net change in the median $\Rorbit$ rather than diffusion about the birth $\Rorbit$, $\sigma(\dRorbit)$.
Thus, one must consider these net changes in the dynamical evolution of a galaxy.

\item \textit{If one selects stars based on $\Rorbit$ today, those born in the pre-disk or early-disk eras (ages $\gtrsim 4 \Gyr$) typically increased in (net) $\Rorbit$ and $j$ since birth, while those born during the late-disk era (ages $\lesssim 4 \Gyr$) typically decreased in $\Rorbit$ ($\lesssim 0.5 \kpc$) and $j$ since birth} (Figures~\ref{fig:dj_v_age_rnow}, \ref{fig:dr_v_rnow}, \ref{fig:dr_v_age_rnow}).
The net positive $\dRorbit$ for older stars is a consequence of a galaxy's inside-out radial growth (older stars formed when the galaxy was smaller) combined with feedback-driven breathing modes that increased $\Rorbit$ for stars at early cosmic times.

\item \textit{If instead one selects stars at a given $\Rorbit$ ($\gtrsim 4 \kpc$) at birth, they typically decreased in $\Rorbit$ and $j$ since birth at almost all most ages} (Figures~\ref{fig:dj_v_age_rform}, \ref{fig:dr_v_age_rform}).
The difference between selecting stars via $\Rorbitnow$ and $\Rorbitform$ (for sufficiently old stars) arises because the stellar density profile declines with radius: more stars at a given $\Rorbitnow$ redistributed outward from smaller radii than redistributed inward from larger radii.

\item \textit{The scatter in radial change, $\sigma(\dRorbit)$, a common metric of radial redistribution (or `radial migration'), increases with age, but only up to $\approx 3 \Gyr$.
It saturates at $\sigma(\dRorbit) \approx 2 \kpc$ at ages $\gtrsim3\Gyr$} (Figure~\ref{fig:dr_v_age_rnow}).

\item \textit{$\dRorbit$ increases with $\Rorbit$ today for all but the youngest stars, while $\sigma(\dRorbit)$ increases with $\Rorbit$ today for all but the oldest stars} (Figure~\ref{fig:dr_v_age_rnow}).
\end{itemize}

\textbf{Correlations of radial redistribution with galaxy properties:}

\begin{itemize}
\item \textit{Galaxies with larger $R^\star_{90}$ or $M^\star_{90}$ today experienced more radial redistribution in terms of $\sigma(\dRorbit)$.
However, we find no significant correlation between the age of disk onset, when $\vphi / \sigma_{v \rm,3D} > 1$, and the amount of radial redistribution in terms of $\sigma(\dRorbit)$} (Figure~\ref{fig:correlation}, Table~\ref{tab:correlation}).
This implies that, in a cosmological setting, the total $\sigma(\dRorbit)$ that stars experienced does not depend on how old the (thin) disk is.
However, earlier-forming disks experienced less net outward redistribution (in terms of $\dRorbit$) since birth, which may arise because earlier-forming disks had less time for feedback-driven breathing modes to drive outward redistribution.
\end{itemize}

\subsection{Discussion}
\label{sec:discussion}

\subsubsection{Caveats}

Several caveats accompany our results, particularly when extending our findings to the MW.

First, we reemphasize that we examined only the changes of stellar orbits between birth and today, without regard to the details of the dynamical evolution over time.
Furthermore, our results for radial redistribution incorporate the combined effects of both cold torquing and (radial) heating, and we do not attempt to disentangle the relative contributions of each.
Changes in $\Rj$ track changes in $j$, but we did not separately track whether this was `cold' versus `hot' torquing.
In Steel et al. (in prep.), we will quantify cold torquing in these simulations in detail.

Second, our FIRE-2 simulations of MW/M31-mass galaxies represent a fairly random (cosmologically representative) sampling of such galaxies today, at least within isolated or LG-like environments, so our results most directly describe the general behavior of MW/M31-mass galaxies and may not necessarily reproduce the history of the MW specifically.
That said, we examined all of our results separately for Romeo, our earliest-forming galaxy/disk, which formed within a LG-like environment, such that it probably provides the best analog to the MW among our sample.
All trends for radial redistribution in Romeo are broadly similar to our FIRE-2 sample as a whole.

Third, the FIRE-2 simulations that we analyze do not model all of the relevant physical processes that act in galaxies.
Some of this omitted physics, such as magnetohydrodynamics (MHD) and cosmic-ray feedback, most likely does not dramatically alter our results regarding stellar radial redistribution.
FIRE-2 galaxies resimulated with MHD have similar stellar masses, star-formation histories, morphologies, and ISM/CGM structure \citep{Su2017, Hopkins2020b}.
Cosmic-ray feedback can have a significant impact on galaxy-wide properties, reducing the star formation rate at late cosmic times, resulting in stellar masses that are 2-3 times lower today \citep{Chan2022, Hopkins2020b}.
However, as \citep{McCluskey2024} showed, this cosmic-ray feedback does not substantially alter the dynamical state of stellar disks in terms of metrics like $v_\phi / \sigma_{v}$ at fixed stellar mass.

The inclusion of supermassive black holes and active galactic nuclei (AGN) feedback could exert a more significant influence on stellar radial redistribution.
AGN feedback is likely a key regulator of gas content and star formation, especially within the inner regions of galaxies \citep{SilkRees1998, Croton2006, Fabian2012, HeckmanBest2014, Alexander2025}. 
By injecting energy and momentum into the surrounding gas, AGN can suppress central star formation and prevent the growth of overly massive bulges \citep{Hopkins2008, Dubois2013, Dave2019}.
In the absence of this regulation, these FIRE-2 galaxies often have significant star formation in their central regions, possibly leading to elevated velocity dispersions and stronger outflows in gas \citep{Chan2022, Gandhi2022, Wellons2023, Marasco2025, Liu2026}.
Enhanced star formation and turbulence in the innermost regions, given the lack of AGN, might lead to increased radial redistribution, at least in the form of radial heating.
That said, as we discuss below, if the lack of AGN also leads to weaker bars, this instead could lead to less radial redistribution, or at least less cold torquing.
That said, the typical $\sigma(\Delta \Rorbit)$ in our simulations is $\approx 2 \kpc$ at $R \gtrsim 8 \kpc$, so the fraction of stars in these $R$ that originate in the inner galaxy is small.

Another important caveat is that these FIRE-2 galaxies rarely host strong long-lived bars.
Although bars do form in about half of these FIRE-2 simulations, they are generally weaker, more ellipsoidal, and shorter-lived than in the observed galaxies \citep{Ansar2025}.
As \citet{Ansar2025} discussed, the dearth of strong long-lived bars in FIRE-2 may stem from the lack of AGN feedback.
Strong bars are expected to form in dynamically cold baryon-dominated disks with relatively low gas fractions, and the Auriga and TNG50 simulations, both of which include AGN, produce stronger and longer-lived bars \citep{Grand2017, Frankel2022, Khoperskov2024, Fragkoudi2020, Fragkoudi2025, Frosst2025}.
That said, those simulations also use significantly different models for ISM cooling, star formation, and stellar feedback than FIRE-2, making it difficult to disentangle which ingredients are decisive for forming and sustaining bars.

Long-lived bars impact stellar dynamics and radial redistribution, because they persistently induce torques that change stars' angular momenta and/or radial actions \citep{LyndenBellKalnajs1972, Athanassoula2003, SellwoodWilkinson1993, Saha2010, Debattista2019}.
In some respects, bars are just additional non-axisymmetric heating agents within the disk, acting to increase orbital eccentricities and thus radial redistribution from blurring/diffusion around birth radii. 
Simulations find that bar strength is well correlated with the degree of heating, particularly for young stars \citep{Grand2016a}.
Therefore, our FIRE-2 simulations may \textit{underestimate} the amount of radial redistribution, relative to strongly barred galaxies.
Furthermore, bars can also induce coherent, disk-wide rearrangements of $j$ via cold torquing.
The formation and subsequent slowdown of a bar can induce substantial coherent outward redistribution, particularly during its early evolutionary phase \citep{Khoperskov2020, Chiba2021, Baba2025}.
Recent work suggests that this bar-driven redistribution may account for the majority of radial redistribution in the MW, inside the outer Lindblad resonance \citep{Haywood2024}. 
In turn, our simulations may underestimate angular momentum changes compared to strongly barred galaxies.

That said, it is not obvious that simulations with weaker bars necessarily experience less radial redistribution overall, because the processes that inhibit bar formation may themselves drive an amount of radial redistribution similar to a bar.
Thus, while it is clear that a bar can influence dynamics within a galaxy, it is less clear exactly how the lack of strong bars in FIRE-2 impacts our results, without a better understanding of the origin of the lack of strong bars in FIRE-2.

\citet{Okalidis2022} compared radial redistribution in strongly-barred and weakly- or non-barred galaxies in the Auriga simulations, finding that while both samples showed similar trends for changes in $\Rphys$, strongly-barred galaxies showed enhanced redistribution in $\Rj$.
Specifically, compared to weakly- and non-barred galaxies, strongly-barred galaxies have higher values for $\sigma(\Delta \Rj)$.
However, both samples exhibit net inward redistribution at all $R$. 

Finally, stars in these FIRE-2 simulations typically have hotter kinematics, in terms of larger $\sigma_{v} / v_\phi$, than those observed in the MW \citep{McCluskey2025}, at least for stars older than a few $100 \Myr$.
\citealt{McCluskey2025} also showed that the MW is dynamically colder than nearly all nearby observed disk galaxies of similar mass.
This suggests that in FIRE-2, stars may experience more radial redistribution from heating than in the MW.
That said, the dynamical state of stars also regulates how strongly they couple to perturbations, with much theoretical work finding that resonance-driven torquing, including `cold torquing' induced by the corotation resonance of spiral arms, is most effective for dynamically cold populations \citep{Solway2012, VeraCiro2014, Halle2015, Daniel2018, Mikkola2020}. 
Hotter stars, in contrast, are less able to `keep station' with the perturbing pattern and are thus less susceptible to cold torquing \citep{SellwoodBinney2002}.
In turn, stars in FIRE-2 simulations may experience weaker changes in $j$ and, importantly, less radial redistribution from cold torquing, compared to stars in the MW, as we will explore in Steel et al. (in prep.).

\subsubsection{Radial redistribution is sub-diffusive and can saturate}

One of the most important findings of our work is that the amount of radial redistribution, as quantified by $\sigma(\Delta \Rorbit)$, does \textit{not} continuously increase with age, but saturates for stars older than $\approx 4 \Gyr$, which also agrees with the inference for the MW from \citet{Ratcliffe2025a}.
This contrasts with a common expectation that radial redistribution follows a diffusive process, increasing as a population ages as $\tau^{0.5}$, where $\tau$ is the stellar age \citep[for example][]{Frankel2020}.

Our results also broadly agree with multiple recent cosmological zoom-in simulations, in that radial redistribution increases more gradually with age.
\citet{Lu2022} examined the strength of radial redistribution in terms of $\sigma(\Delta \Rphys)$ as a function of stellar age for $\approx 100$ galaxies in the NIHAO zoom-in simulations \citep{Buck2020nihao}, finding that $\sigma(\Delta \Rphys) / R_{\rm d} \approx 0.59 \times \tau^{0.32}$, where $R_{\rm d}$ is a galaxy's disk scale radius.
Thus, radial redistribution is a sub-diffusive process in the NIHAO simulations, increasing more slowly with age than $\tau^{0.5}$.

Similarly, \citet{Okalidis2022} parameterized radial redistribution in the Auriga zoom-in simulations, fitting diffusion relations for $\sigma(\Delta \Rphys)$ and $\sigma(\Delta \Rj)$ at 3 radial ranges in each of their simulated galaxies.
At solar neighborhood-like radii, they found that radial redistribution increases with age typically as $\tau^{0.35 - 0.40}$.

\citet{Dubay2024} examined radial redistribution in the h277 cosmological zoom-in simulation \citep{Christensen2012}, finding that it is well described by $\sigma(\Delta \Rphys) = \sigma_{RM8} (\tau / 8 \Gyr)^{0.33} (R_{\rm form} / 8 \kpc) ^{0.61}$, with $\sigma_{RM8} = 2.68 \kpc$ corresponding to the redistribution strength of an 8 Gyr-old population at $R_{\rm form} = 8 \kpc$.
Their age dependence, $\tau^{0.33}$, is again sub-diffusive, broadly mirroring the behavior we find. 
h277 provides a particularly compelling comparison because it remains, to our knowledge, the only cosmological zoom-in simulation that reproduces the cold stellar kinematics of the MW, given h277's early-forming disk with a quiescent merger history \citep{Bird2021}.
See \citet{McCluskey2025} for a broader discussion on stellar kinematics in cosmological simulations and h277's unique behavior.

Although modern cosmological zoom-in simulations -- including FIRE-2, NIHAO, Auriga, and h277 -- persistently find that radial redistribution is sub-diffusive, increasingly with age more slowly than $\tau^{0.5}$, theoretical models attempting to model and/or recover radial redistribution in the MW routinely parameterize radial redistribution via a diffusion process \citep[for example][]{SandersBinney2016, Frankel2020, Sharma2021dynamics}.
Indeed, the monotonic increase in $\sigma(\Delta R)$ reported by \citet{Frankel2020} (and shown in Figure~\ref{fig:literature_compare_mw} reflects their model's fit to such a form.

Our results, combined with numerous other cosmological simulations, motivate substantially more flexibility in modeling assumptions about radial redistribution of stars in the MW, and all disk galaxies, in terms of $\sigma(\Delta R)$, including weaker dependence on age and a possible saturation (on average) for stars older than $\approx 4 \Gyr$ at $\sigma(\Delta R) \approx 2 - 3 \kpc$.

\section*{Acknowledgements}

We performed this work using the \textsc{GizmoAnalysis} package \citep[][]{Hopkins2015}, the Astropy package \citep[][]{Astropy2013, Astropy2018}, as well as libraries from Numpy \citep[][]{Numpy2020}, SciPy \citep[][]{Scipy2020}, and Matplotlib \citep[][]{Matplotlib2007}.
We also used the online tool WebPlotDigitizer \citep[][]{Webplotdigitizer2022}.

MAB and AW received support from: NSF via CAREER award AST-2045928 and grant AST-2107772.
FM and CS received support from NASA FINESST award 80NSSC24K1484.

We generated simulations using: XSEDE, supported by NSF grant ACI-1548562; Blue Waters, supported by the NSF; Frontera allocations AST21010 and AST20016, supported by the NSF and TACC; Pleiades, via the NASA HEC program through the NAS Division at Ames Research Center.

The FIRE-2 simulations are publicly available \citep{Wetzel2023, Wetzel2025} at \url{http://flathub.flatironinstitute.org/fire}.
Additional FIRE simulation data is available at \url{https://fire.northwestern.edu/data}.
A public version of the GIZMO code is available at \url{http://www.tapir.caltech.edu/~phopkins/Site/GIZMO.html}.

\bibliographystyle{mnras}
\bibliography{radial_redistribution}{}

\appendix


\section{A. Effect of Orbital Radius on Metallicity Profiles}
\label{sec:abundance_profile}

\renewcommand{\thefigure}{A\arabic{figure}}
\setcounter{figure}{0}

\begin{figure}
\centering
\includegraphics[width = \linewidth]{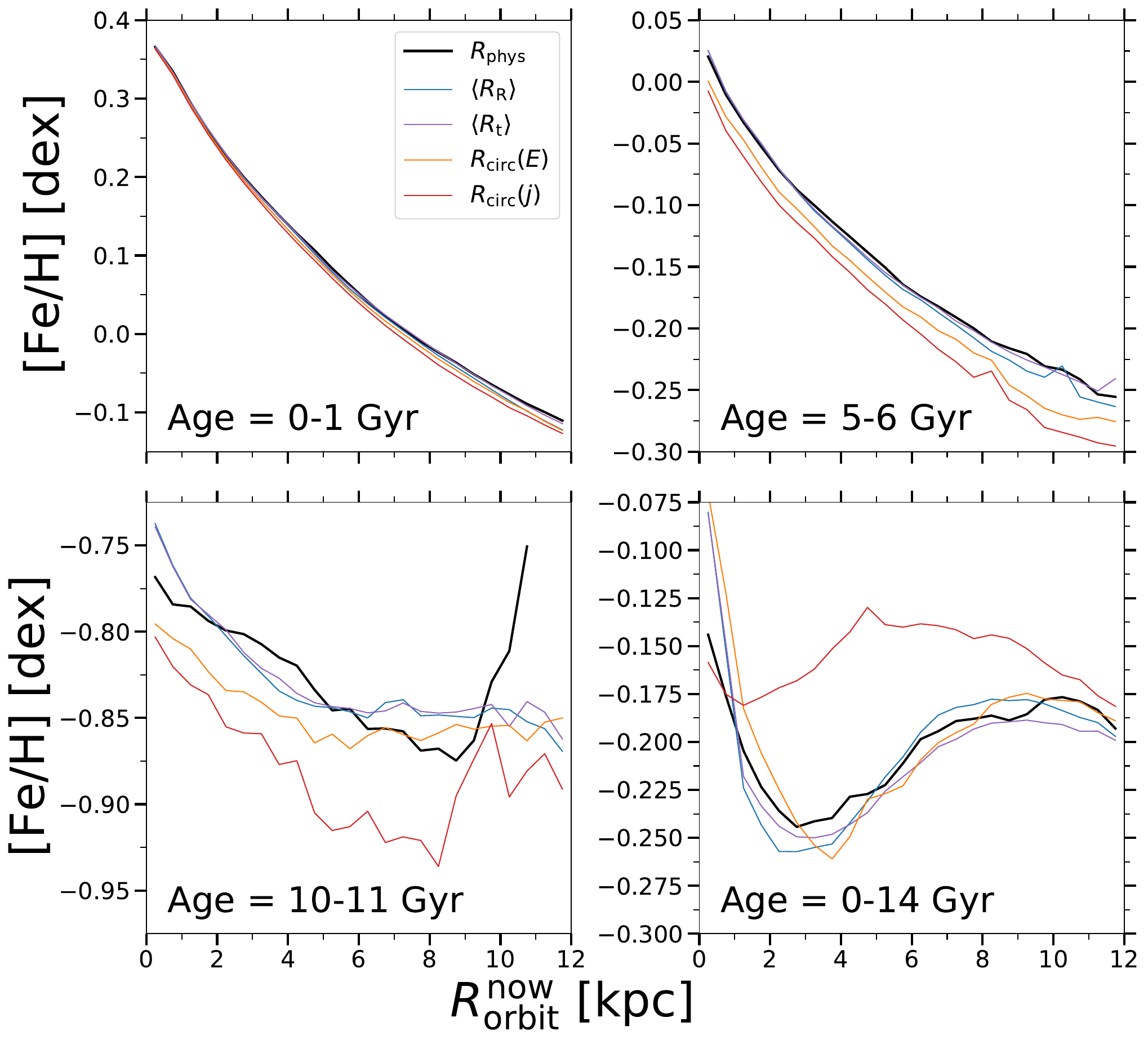}
\caption{
Stellar [Fe/H] versus current orbital radius, $\Rorbitnow$, in bins of stellar age, averaged across our galaxies.
Lines compare our 5 different metrics of orbital radius.
The bottom right panel shows all stars, regardless of age, today.
The choice of orbital radius metric moderately affects the normalization of the abundance profile, more so for older stars, with $\Rj$ showing the lowest normalization and $\Rphys$ showing the highest normalization.
However, the radial gradients are all consistent within $\lesssim 0.005$~dex kpc$^{-1}$ at any age.
The [Fe/H] profile for all stars also agrees across the radius metrics, except for $\Rj$, which is systematically larger at most $R$.
\textit{We conclude that the choice of orbital radius metric only weakly affects the metallicity radial gradient.}
}
\label{fig:abundance_profile}
\end{figure}

To explore further the effect of the metric for orbital radius, $\Rorbit$, on galaxy properties, Figure~\ref{fig:abundance_profile} shows the average [Fe/H] versus present-day orbital radius, $\Rorbitnow$, for stars in bins of age, using the 5 metrics of $\Rorbit$.
We show results for mono-age populations sampling the three different eras of disk formation (pre-disk, early-disk, and late-disk), specified in Section~\ref{sec:transition_epochs}.

The primary effect of varying the metric of $\Rorbit$ is an offset to the normalization of the abundance profile at a given $\Rorbit$.
This is a weak effect for the young stars, which are all on near-circular orbits.
The difference is strongest for the oldest stars, which tend to be on the most eccentric orbits, so the metrics of $\Rorbit$ are most different (see Figure~\ref{fig:radius_metric}).

However, we find that the choice of radius metric has little effect on the \textit{gradient}.
We measure the strength of the radial abundance in [Fe/H] for each definition of orbital radius for stars in $1 \Gyr$ bins of age (not shown), and we find that the gradient is always consistent within $\lesssim 0.005$~dex kpc$^{-1}$ at any age, at least at sufficiently small $R$ where a single slope describes the gradient well.
This is a small effect relative to the variations in radial gradients across these galaxies, from $-0.24$ to $-0.05 \text{dex kpc}^{-1}$ for young stars today \citep{Bellardini2022}.

This result is generally consistent with the observational analysis of the MW in \citet{Lian2022} (their Figures 3 and A2).
They measured [Fe/H] of stars as a function of stellar age in different bins of $R$ (analogous to our $\Rphys$) and `guiding center radius' (analogous to our $\Rj$).
The radial profiles of their sample are qualitatively consistent regardless of their definition of $R$, but their profiles show a normalization offset.
The notable difference is that their abundance profile measured via guiding radius is not consistently lower [Fe/H] at a given $R$, unlike our abundance profile measured with $\Rj$.

Figure~\ref{fig:abundance_profile} (bottom right) also shows the abundance profile for all stars today, rather than for mono-age populations.
All metrics of $\Rorbit$ are quantitatively consistent (beyond $\approx 1 \kpc$) except for $\Rj$.
The abundance profile for all stars measured via $\Rj$ is systematically larger for all radii $\gtrsim 1 \kpc$.
This is likely driven by the negative radial gradient in age within these galaxies \citep[for example][]{Ma2017b}, a result of inside-out radial growth \citep{Graf2025b}, in conjunction with $\Rj$ being a systematically smaller measure of radius than the other radial metrics.
Figure~\ref{fig:abundance_profile} (bottom right) also shows a `break' $R$, where the galaxies transition from a negative gradient in [Fe/H] in the inner galaxy to a positive/flat gradient in the outer galaxy \citep[see also][]{Graf2025a}.
\citet{Lian2023} found a similar break $R$ in the light-weighted stellar [Fe/H] profile of the MW.
Our results suggest that using $\Rj$ leads to a systematically smaller break $R$.

\section{B. Effect of radial selection across the galaxy}
\label{sec:radial_selection}

\renewcommand{\thefigure}{B\arabic{figure}}
\setcounter{figure}{0}

\begin{figure*}
\centering
\includegraphics[width = \linewidth]{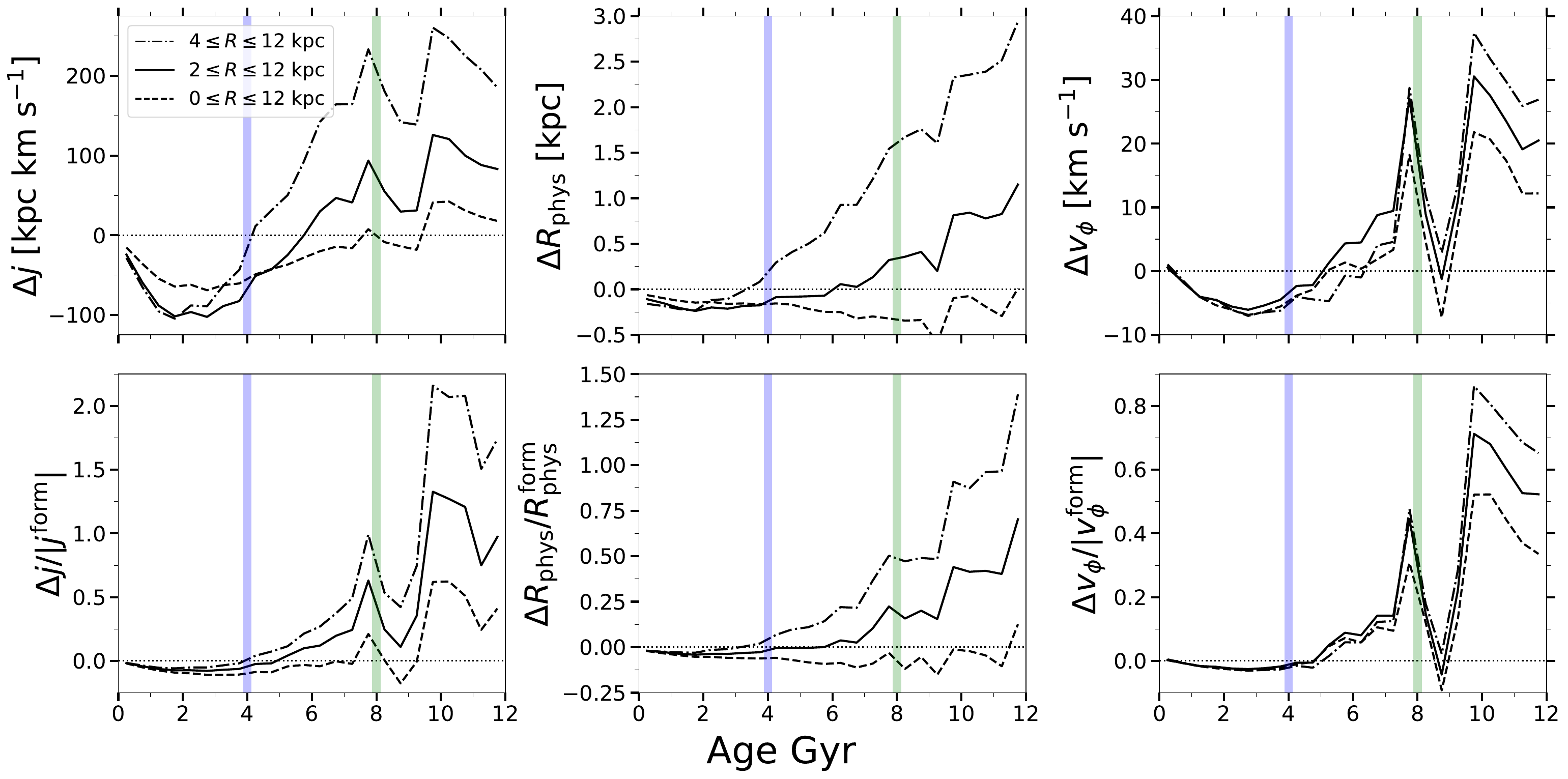}
\caption{
Same as Figure~\ref{fig:jrv_v_age}, but selecting stars across different ranges of $\Rphys$ today.
Solid lines show our fiducial selection.
Because of the declining density profile of stars, the choice of minimum $\Rphys$ significantly can change the population of stars included, so selecting stars just in the `disk' region ($R \gtrsim 2 - 4 \kpc$) versus including stars in the `bulge' region can affect the picture of the galaxy overall.
In general, the radial selection has a weak effect for younger stars, but the increases in angular momentum, radius, and azimuthal velocity for older stars are weaker when including stars at smaller $\Rphys$ today.
}
\label{fig:jrv_v_age_radius}
\end{figure*}

Figures~\ref{fig:j_v_r} and \ref{fig:jrv_v_age} showed trends for all stars across the disk, within a range $R^{\rm now}_{\rm phys} = 2 - 12 \kpc$.
This upper limit approximately corresponds to the typical $R^\star_{90}$ of these galaxies \citep[see][Table 1]{Bellardini2021}, though the lower limit is more ambiguous.
We use $R^{\rm now}_{\rm phys} > 2 \kpc$, based on where stars generally form on circular orbits in these simulations at $z \approx 0$.
Here we test the dependence of our results on that choice, by also using $R^{\rm now}_{\rm phys} > 0 \kpc$ and $> 4 \kpc$.

We do not show the effects on Figure~\ref{fig:j_v_r} (right), because we find essentially no change when varying the selection minimum $R^{\rm now}_{\rm phys}$.
Selecting all stars in the disk shifts the amount of redistribution down by an average of $\approx 5$\% and selecting stars at $4 - 12 \kpc$ shifts the trend up by $\approx 3$\%.
The shape is unchanged.
The variation is most noticeable for stars older than $\approx 8 \Gyr$, because they have the strongest radial dependence to their orbital circularity.

Figure~\ref{fig:jrv_v_age_radius} recreates Figure~\ref{fig:jrv_v_age} using different selection minimum $R^{\rm now}_{\rm phys}$.
Most trends remain qualitatively similar, especially $\Delta \vphi$.
The biggest differences are for the oldest stars.
Specifically, when selecting all stars across $R^{\rm now}_{\rm phys} = 0 - 12 \kpc$, $\Delta j$ and $\Delta \Rphys$ are near 0 or even negative, even at old ages. This is because stars in the inner galaxy today have experienced little or negative $\Delta j$ and $\Delta \Rphys$, and the declining density profile of these galaxies means that a significant fraction of stars today are at $R^{\rm now}_{\rm phys} < 2 \kpc$, which shifts the disk-wide average.
This, coupled with systematically older stars near the galactic center, leads to a significant decrease in outward redistribution for these effects for older stars.
We find the opposite effect in selecting stars at larger $R^{\rm now}_{\rm phys} = 4 - 12 \kpc$.

This test reinforces that the trends in radial redistribution depend on the radial range one uses to select stars, and `galaxy-wide' trends can be substantially different depending on the exact radial range of choice.
This is why we examine most trends in this work for stars at a given $R^{\rm now}$ or $R^{\rm form}$.

\section{C. Selecting stars by birth radius}
\label{sec:birth_radius}

\renewcommand{\thefigure}{C\arabic{figure}}
\setcounter{figure}{0}

\begin{figure*}
\centering
\includegraphics[width = \linewidth]{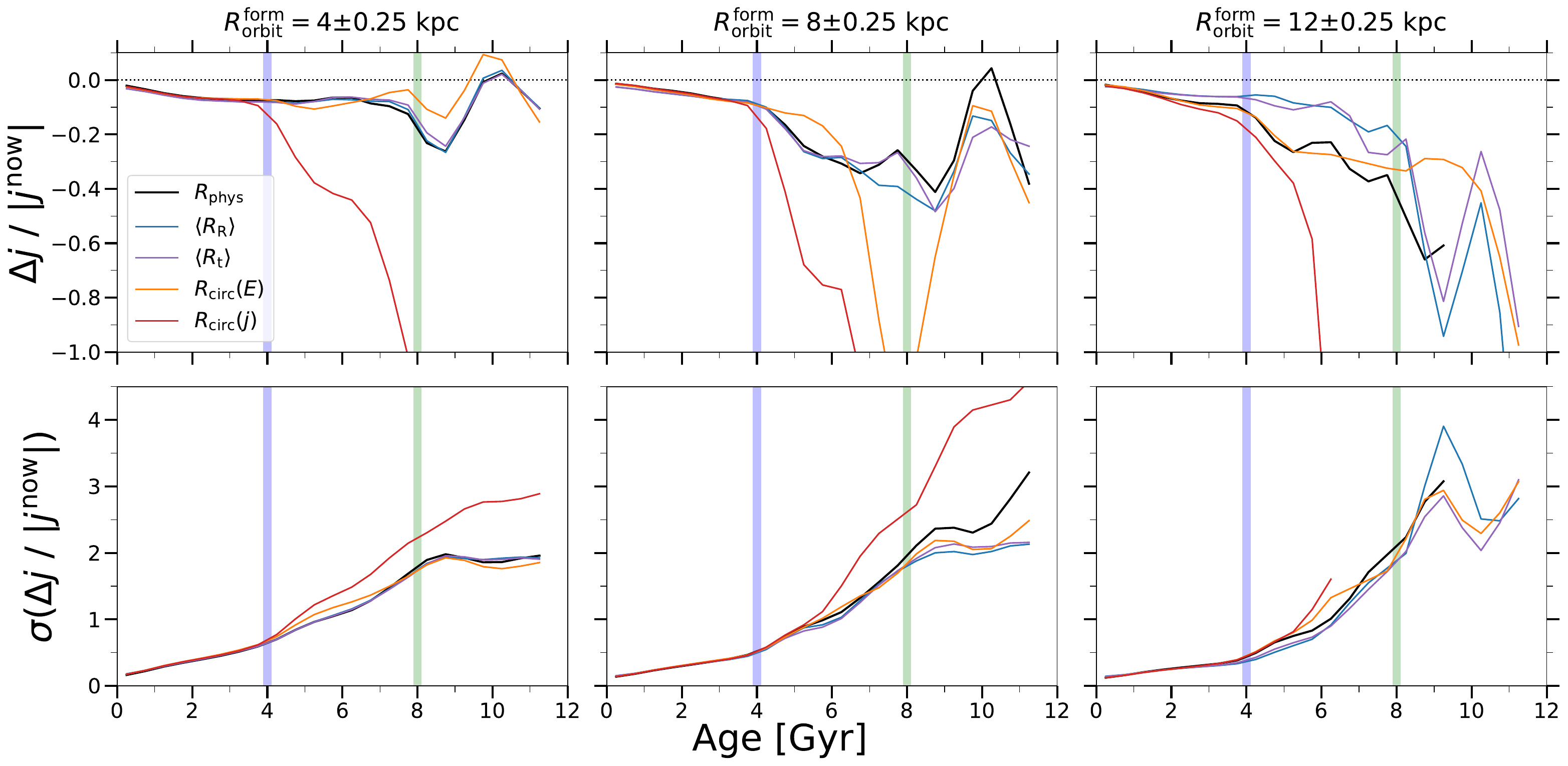}
\caption{
Same as Figure~\ref{fig:dj_v_age_rnow}, but selecting stars based on orbital radius \textit{at birth}, instead of today.
Now, most stars that formed at a given $\Rorbit$ have lost $j$ since birth.
The gain in $j$ in Figure~\ref{fig:dj_v_age_rnow} when selecting (sufficiently old) stars by $\Rorbitnow$ arises because of the declining density profile of the galaxy: more stars at a given $\Rorbitnow$ redistributed outward from smaller $j$ than redistributed inward from larger $j$.
This fractional decrease in $j$ is approximately constant with age for stars that formed in the inner galaxy, but it grows with age for stars that formed at larger radii.
Similar to Figure~\ref{fig:dj_v_age_rnow}, $\sigma$ (bottom) increases with age.
}
\label{fig:dj_v_age_rform}
\end{figure*}

\begin{figure*}
\centering
\includegraphics[width = \linewidth]{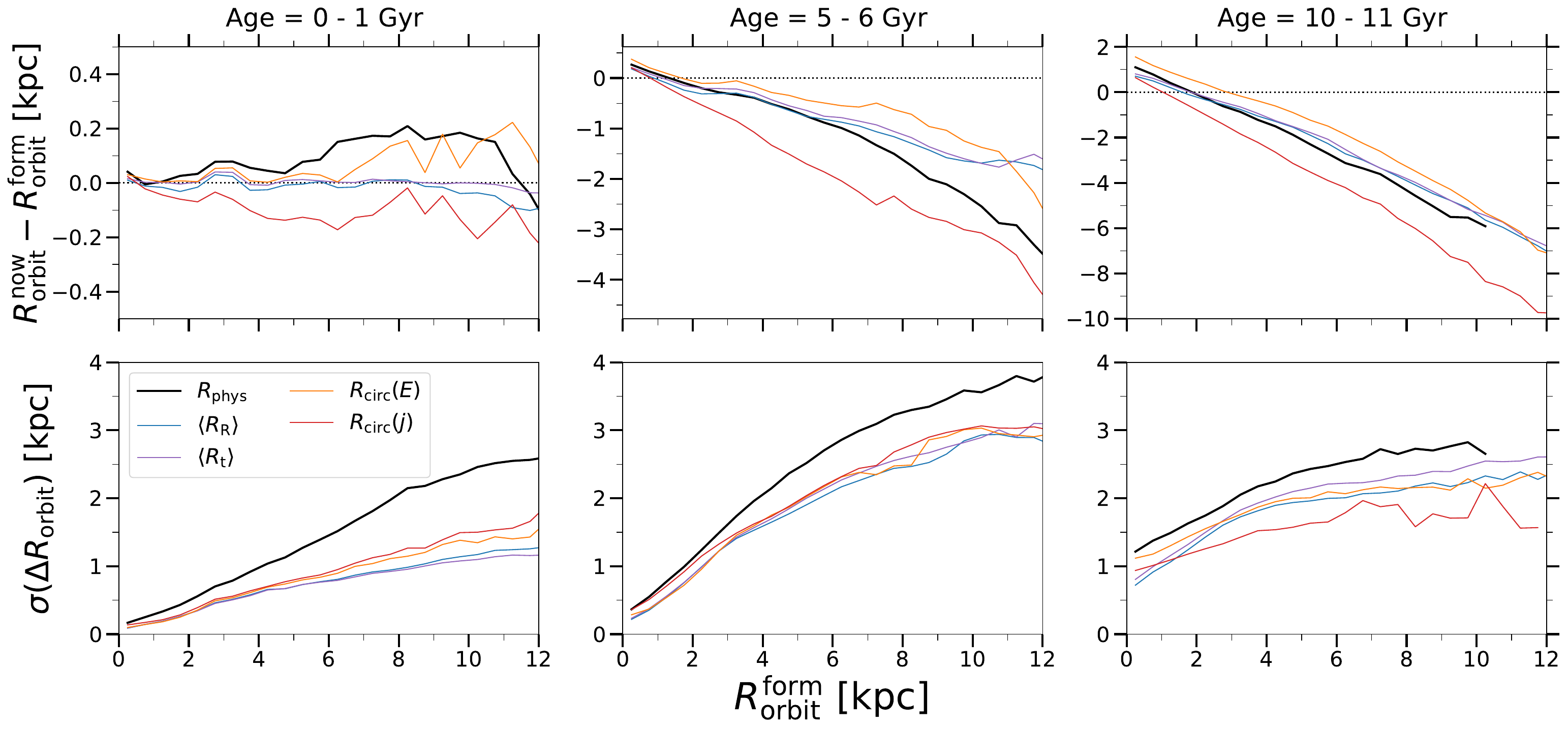}
\vspace{1 mm} \\
\includegraphics[width = \linewidth]{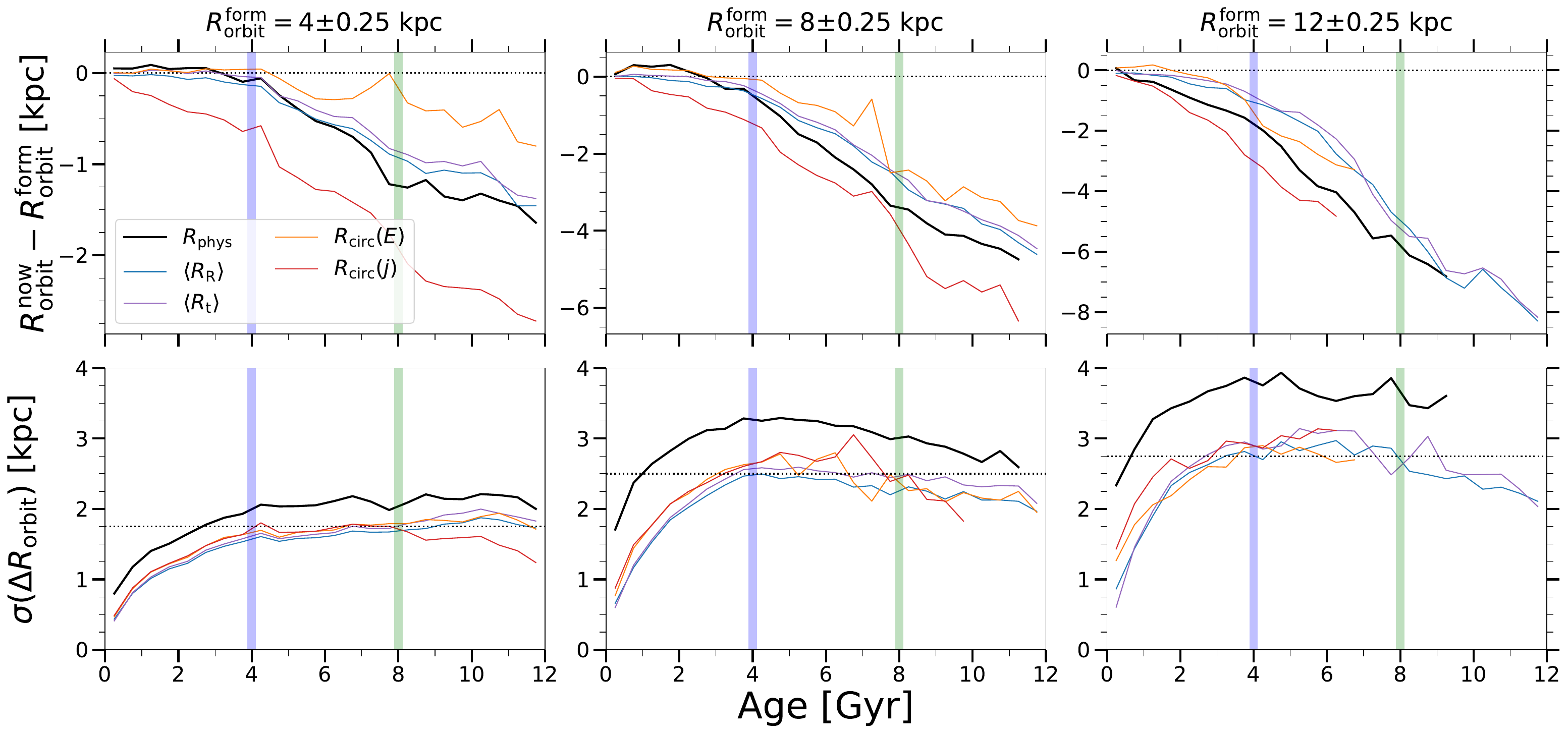}
\caption{
Same as Figures~\ref{fig:dr_v_rnow} and \ref{fig:dr_v_age_rnow}, but selecting stars based on orbital radius \textit{at birth}, instead of today.
Top panels show radial redistribution versus $\Rorbitform$ in bins of stellar age, while bottom panels show the same trends versus stellar age in bins of $\Rorbitform$.
Now, the median $\dRorbit$ is almost always negative, with stronger inward redistribution for older stars and at larger radii.
The outward redistribution in Figures~\ref{fig:dr_v_rnow} and \ref{fig:dr_v_age_rnow}, when selecting (sufficiently old) stars by $\Rorbitnow$, arises because of the declining density profile of the galaxy: more stars at a given $\Rorbitnow$ redistributed outward from smaller $R$ than redistributed inward from larger $R$.
However, $\sigma(\dRorbit)$ here is remarkably similar to that in Figures~\ref{fig:dr_v_rnow} and \ref{fig:dr_v_age_rnow}, regardless of whether one selects stars via $\Rorbitform$ or $\Rorbitnow$.
}
\label{fig:dr_v_age_rform}
\end{figure*}

Previously, we showed trends selecting stars based on their orbital radius today, motivated by observational analyses of the MW and nearby galaxies.
Here, we instead examine trends selecting stars by their orbital radius at birth, $\Rorbitform$.

Similar to Figure~\ref{fig:dj_v_age_rnow}, Figure~\ref{fig:dj_v_age_rform} shows the fractional change in angular momentum, $j$, relative to its value today, as a function of age, but now for stars selected by $\Rorbitform$.
Now, across nearly all ages and $R$, the median $\Delta j < 0$, unlike in Figure~\ref{fig:dj_v_age_rnow}.
Again, this difference mainly reflects the declining stellar density profile of galaxies, such that even if the same number of stars at a given $\Rorbitform$ increased and decreased in $j$ since birth, at a given $\Rorbitnow$, a larger population will have redistributed outward from smaller $j$ than inward from larger $j$.

Again, the fractional difference in $\Delta j$ is smaller for younger stars, consistent with disk settling and younger stars forming on more circular orbits.
There is little to no dependence on age for stars born in the inner disk, but the dependence on age is strong for stars born at larger radii.

Stars selected via $\Rj$ at birth experienced a larger fractional decrease in $j$.
This is because stars at a given $\Rj$ have the maximum possible $j$, by definition, so at a given $\Rorbitform$ they had more $j$ to lose.
By contrast, for the other metrics of $\Rorbit$, stars on eccentric orbits could have large $\Rorbitform$ even if they had small $j$, so on average they had less $j$ to lose.

Figure~\ref{fig:dj_v_age_rform} (bottom) shows that the trends for $\sigma(\Delta j / |j^{\rm now}|)$ for stars selected by $\Rorbitform$ are broadly consistent across all metrics.
This contrasts with Figure~\ref{fig:dj_v_age_rnow} (bottom), where stars selected via $\Rj$ have systematically lower scatter than the other metrics.
This is because we divide by $|j^{\rm now}|$: in Figure~\ref{fig:dj_v_age_rnow} $|j^{\rm now}|$ correlates exactly with $R^{\rm now}_{\rm circ}(j)$ (by definition), where here $|j^{\rm now}|$ correlates more weakly with $R^{\rm form}_{\rm circ}(j)$ , so this adds more scatter.

Figure~\ref{fig:dr_v_age_rform} is the same as Figures~\ref{fig:dr_v_rnow} and \ref{fig:dr_v_age_rnow} except we select stars via $\Rorbitform$.
Similar to the trends for $j$, stars selected by $\Rorbitnow$ experienced $\dRorbit > 0$, while here stars selected by $\Rorbitform$ generally experienced $\dRorbit < 0$.
Thus, stars typically redistribute inward from their birth $R$, so the positive redistribution of stars when selecting them by $\Rorbitnow$ again stems from the declining density profile of galaxies.

Figure~\ref{fig:dr_v_age_rform} also shows $\sigma(\dRorbit)$, which shows similar trends as for selecting stars via $\Rorbitnow$.
In particular, the value of $\sigma(\dRorbit)$ at a given age or $\Rorbit$ is remarkably similar regardless of whether one selects stars based on $\Rorbitnow$ or $\Rorbitform$.

Figure~\ref{fig:dr_v_age_rform} qualitatively agrees with \citet{Okalidis2022} (their Figure 6).
They found that, at a given birth $R$, the peak present-day $R$ of stars decreases with age, that is, older stars redistribute inward more than younger stars.

\section{D. Selecting stars by instantaneous physical radius}
\label{sec:rphys}

\renewcommand{\thefigure}{D\arabic{figure}}
\setcounter{figure}{0}

\begin{figure*}
\centering
\includegraphics[width = \linewidth]{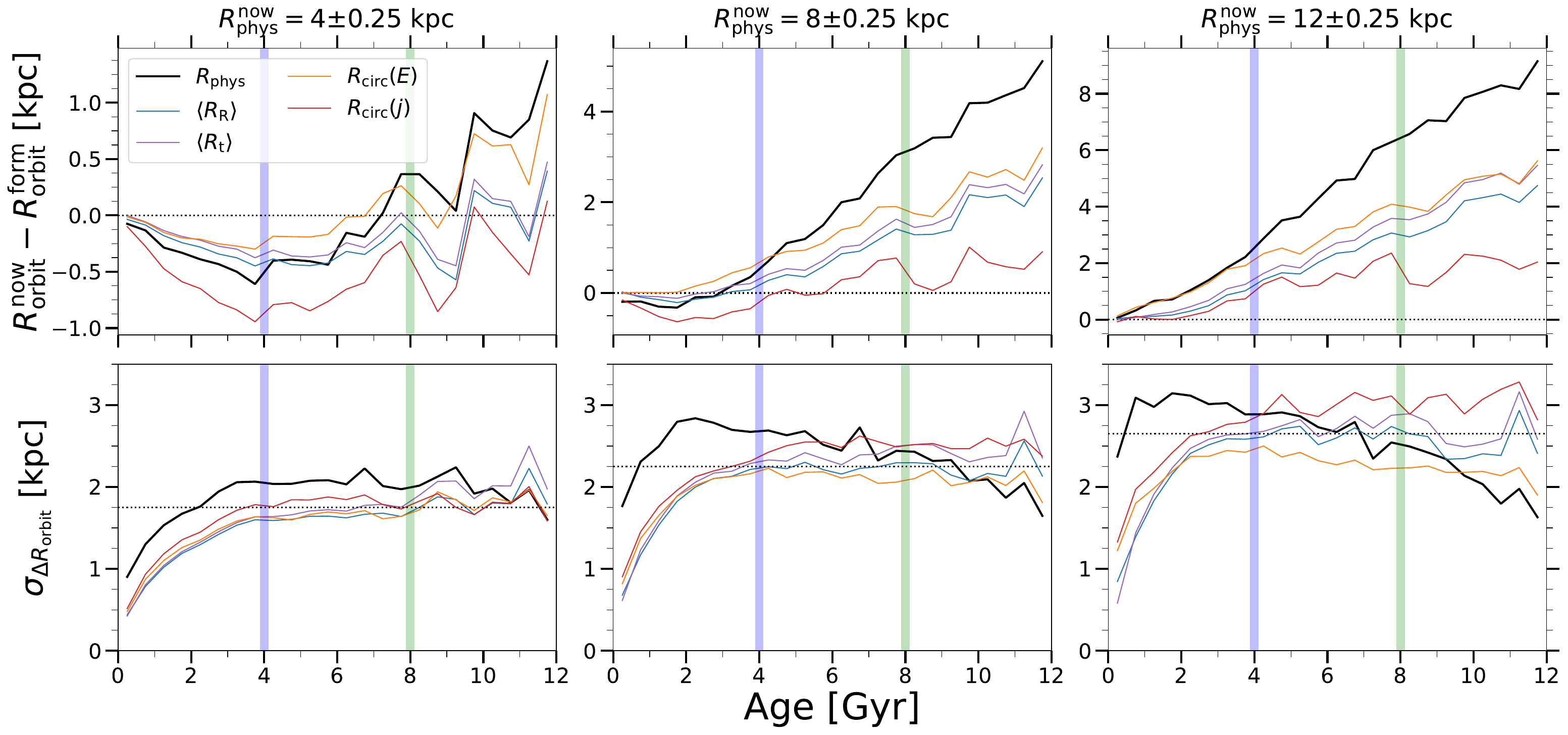}
\caption{
Same as Figure~\ref{fig:dr_v_age_rnow}, using different metrics of orbital radius to measure radial redistribution, but here in all cases we select stars via $\Rphys$ today.
Selecting stars using a different radius metric than for measuring radial redistribution leads to significantly less agreement between the radius metrics, especially in terms of $\dRorbit$.
\textit{This motivates the importance of being self-consistent in how one selects stars and how one measures radial redistribution.}
}
\label{fig:dr_v_age_rphys}
\end{figure*}

Throughout, we have shown that the results for radial redistribution, in terms of both $\dRorbit$ and $\sigma(\dRorbit)$, are remarkably similar, regardless of the metric for $\Rorbit$ used, as long as one self-consistently uses the same $\Rorbit$ to select stars and measure redistribution.
Here we show that using different metrics of $\Rorbit$  when selecting stars versus measuring redistribution leads to less similar results, and with larger scatter.

Figure~\ref{fig:dr_v_age_rphys} is the same as Figure~\ref{fig:dr_v_age_rnow}, except in all cases we select stars via $\Rphysnow$ (in bins of width $0.5 \kpc$).
While the trends in Figure~\ref{fig:dr_v_age_rphys} (top) are qualitatively consistent with Figure~\ref{fig:dr_v_age_rnow} (top), they are quantitatively different.
In particular, there is much less similarity among the metrics, and $\dRorbit$ is more likely to be negative at small $\Rphysnow$.
$\dRorbit$ tends to be most positive for $\Rphys$ and smallest (or most negative) for $\Rj$.

$\sigma(\dRorbit)$ is systematically larger selecting stars via $\Rphysnow$ than selecting via $\Rorbit$, espeically for for $\Rj$, as in Figure~\ref{fig:dr_v_age_rnow}.
That said, the increase is typically $\lesssim 1 \kpc$.

\end{document}